\newcommand{\timestamp}{Time-stamp: "2004-08-07 20:06:56 daboul"}
\documentclass[11pt,twoside,final,onecolumn,a4paper]{article}
\usepackage[height=23cm,inner=2.5cm,outer=2cm]{geometry}
\usepackage{psfrag}
\usepackage[dvips]{graphicx} %
\usepackage[notcite,notref]{showkeys} %
\usepackage{amsmath}
\usepackage{amssymb} %
\usepackage{cite} %
\usepackage{ifthen}
\usepackage{longtable}
\usepackage{multirow}
\usepackage{layout}
\newboolean{bo:thesis}
\setboolean{bo:thesis}{false}
\newcommand{\inthesis}[2]{
  \ifthenelse{\boolean{bo:thesis}}{#1}{#2}
}
\newboolean{bo:aa}
\setboolean{bo:aa}{true} %
\ifthenelse{\boolean{bo:aa}}{\linespread{1.6}}{}

\newcommand{\be}{\begin{equation}}
\newcommand{\ee}{\end{equation}}

\newcommand{\kb}{k_\mathrm{B}}
\newcommand{\ka}{\kappa}
\newcommand{\kc}{k}
\newcommand{\kx}{k}
\newcommand{\ku}{u}
\newcommand{\kex}{x}
\def\Tr{\mathop{\mbox{Tr}}}
\newcommand{\EA}{\chi_{\mbox{}_\mathrm{EA}}}
\newcommand{\qea}{q_{\mbox{}_\mathrm{EA}}}
\newcommand{\sts}{}
\newcommand{\multsp}{\,}
\newcommand{\et}{{\it et al.\/}}
\newcommand{\dlog}{Dlog-Pad\'e}

\newcommand{\Jcon}{J} %
\newcommand{\rvar}{z} %
\newcommand{\pn}{x_{\mathrm{n}}}
\newcommand{\pc}{x_{\mathrm{c}}}
\newcommand{\pq}{x_{\mathrm{c}}}

\begin{document}
\renewcommand{\timestamp}{Time-stamp: "2004-08-07 20:03:05 daboul"}
\inthesis{
}{
  \title{Test of Universality in the Ising Spin Glass Using \\
    High Temperature Graph Expansion.}
\author{
  Daniel Daboul$^a$,
  Iksoo Chang$^b$ and
  Amnon Aharony$^a$\\
  \em $^a$School of Physics and Astronomy,\\ \em Raymond and
  Beverly Sackler Faculty of Exact Sciences,\\ \em Tel Aviv University,
  69978 Tel Aviv, Israel\\
  $^b$\em Department of Physics,\\
  \em Pusan National University, Pusan 609-735,
  Korea
}
\date{August 7, 2004}
\maketitle\sts
}
\inthesis{
}{
\begin{abstract}
} We calculate high-temperature graph expansions for the Ising spin
glass model with 4 symmetric random distribution functions for its
nearest neighbor interaction constants $J_{ij}$.  Series for the
Edwards-Anderson susceptibility $\EA$ are obtained to order 13 in the
expansion variable $(\Jcon/(k_\mathrm{B}T))^2$ for the general
$d$-dimensional hyper-cubic lattice, where the parameter $\Jcon$
determines the width of the distributions.  We explain in detail how
the expansions are calculated.  The analysis, using the \dlog\ 
approximation and the techniques known as M1 and M2, leads to
estimates for the critical threshold $
(\Jcon/(k_\mathrm{B}T_\mathrm{c}))^2$ and for the critical exponent
$\gamma$ in dimensions 4, 5, 7 and 8 for all the distribution
functions.  In each dimension the values for $\gamma$ agree, within
their uncertainty margins, with a common value for the different
distributions, thus confirming universality.

\inthesis{}{
and other random models
\end{abstract}
}
\newcommand{\nn}{{\langle ij\rangle}} %
\inthesis{
  \newcommand{\art}{chapter}}{
  \newcommand{\art}{article}}
\section{Introduction}

In 1975 Edwards and Anderson (EA)~\cite{EdwardsA75} introduced a model
for the theoretical study of spin glasses (SG)~\cite{binderyoung,
  fisherhertz}, which has started modern spin glass theory and has
been of continued interest until today. \inthesis{It is a problem of
  classical statistical mechanics of which we study the Ising
  case:}{Here we discuss the classical Ising case:} The magnetic
moments are represented by `spin' variables $\{s_i,i=1, 2,\ldots,
N\},$ which can assume the values $s_i=\pm1$ and are located on the
sites $\{i\}$ of the $d$-dimensional hyper-cubic lattice.  During our
calculations we use a finite number of lattice sites $N$ but
eventually we are interested in the thermodynamic limit,
$N\rightarrow \infty$.  The spins' interaction is governed by the
Hamiltonian
\begin{equation}\label{eq:hamsg}
\mathcal{H}_{\{J_{ij}\}}(\{s_i\})=-\sum_\nn J_{ij}\,
s_i s_j \
-h_0\sum_{i=1}^N  s_i\,,
\end{equation}
where $\sum_\nn$ denotes the sum over all pairs of nearest neighbor
lattice sites $\nn$, which we also call the lattice bonds, and the
spin interaction constants $J_{ij}$ are chosen at random from a
symmetric probability distribution, which is the same for all bonds.
The external magnetic field $h_0$ is needed to define thermodynamic
quantities as derivatives with respect to it, but apart from that, we
concentrate on the case $h_0=0$. The Hamiltonian's index $\{J_{ij}\}$
indicates that we deal with {\em quenched disorder}, i.e.\ the {\em
  thermodynamic average} for any observable is performed for a fixed
set of coupling constants $\{J_{ij}\}$.  The {\em configurational
  average} of measurable thermodynamic quantities, over the random
variables, is performed subsequently. For self averaging quantities
this leads to expressions of what could be measured in experiments. We
denote the thermodynamic average of any observable $A(\{s_i\})$ by
$\langle A \rangle_T$, and the configurational average of any function
$X(\{J_{ij}\})$ by $\left[X \right]_\mathrm{R}$.

The EA model neglects the details of the microscopic interaction
between the spins, but exhibits the two essential ingredients that
lead to the interesting features of spin glasses: {\em Quenched
  disorder} and {\em frustration}. Since little has been proved
exactly for short ranged spin glass models, we assume what today is
generally accepted, based on analytical and numerical evidence: Above
the system's lower critical dimension $d_l$, whose value is
controversial but agreed to be between 2 and
3~\cite{Young02,AmorusoMMP03}, it undergoes a continuous transition at
a non-zero critical temperature $T_\mathrm{c}$ to a low temperature
spin glass phase. This phase is characterized by broken spin-flip
symmetry, i.e.\ a non-zero {\em Edwards-Anderson order parameter }
\begin{equation}
  \qea=\frac{1}{N}
  \sum_{i=1}^N
  \left[\langle s_i \rangle^2_T\right]_\mathrm{R}.
\end{equation}
The upper critical dimension, above which mean field behavior becomes
dominant, is believed to be $d_u=6$ \cite{DominicsKT98,WangY93}.

As the temperature $T$ approaches $T_\mathrm{c}$ from above, we expect
the susceptibility associated with $\qea$, the {\em Edwards-Anderson
  susceptibility},
\begin{equation}\label{eq:EA}
  \EA=%
  \frac{1}{N}
  \sum_{i,j=1}^N
  \left[\langle s_i s_j \rangle^2_T\right]_\mathrm{R}\,,
\end{equation}
to exhibit a power law divergence, $\EA\sim(T_\mathrm{c}-T)^{
  -\gamma}$, characterized by the critical exponent $\gamma$.  In the
present study we use series expansions to investigate this behavior.
Both $\qea$ and $\EA$ are related to configurational averages of
higher order logarithmic derivatives of the partition function
$\left.-\frac{\partial^m \ln Z}{\partial h_0^m}\right|_{h_0=0}$ with
respect to the external magnetic field. Those relations become linear
in the thermodynamic limit~\cite{KleinAAHM91}.

The renormalization group theory~\cite{PytteR79} in dimension $d=6-
\varepsilon$ predicts the universality of $\gamma$ and of other exponents,
related to it by scaling relations. The universality classes should be
set by the dimensionalities of space and of the spin variables, and
not by details of the distribution functions.

Series expansion has been used in the past to study the spin glass
transition \cite{SinghC86,KleinAAHM91,Singh91,HetzelBS93,SinghA96} and
the results support the statements mentioned above. Our renewed
interest in the problem awoke with a series of studies \cite{camp94e,
  camp94b, BernardiC95, BernardiPC96, BernardiC97, BernardiLMCAC98,
  CampbellPMB00} that found, based on computer simulations, that the
critical exponents vary with the probability distribution for the
quenched disorder in the coupling constants $J_{ij}$. This is in clear
violation of universality and not sufficiently explained by theory.

Undoubtedly, many of the enormous complications and features observed
in the study of spin glasses arise from the disorder inherent in these
systems. They gave the model the reputation of being one of the
toughest subjects in computational physics.  Simulations are here
directly impacted by long relaxation times, memory effects,
hysteresis, the rugged energy landscape with many meta-stable states
and the huge parameter space over which to average.

The technique of series expansion comes with two immediate advantages:
The averaging over the randomness can be done {\em exactly}, and the
series can, given the availability of graph data, be obtained in
general dimension. The subsequent analysis is still done in each
dimension separately, but results generally get more reliable with
increasing dimension, while simulations become increasingly expensive
in their computational demands.  The previous series expansion studies
of the Ising spin glass used only the bimodal random distribution of
$J_{ij}=\pm \Jcon$, limiting their use in the comparison with the claims
of violated universality.  In the present study we extend the research
by addressing several other symmetric distribution functions, each
with a variable width determined by the parameter $\Jcon$. We use the
same distributions as Bernardi and Campbell in~\cite{camp94b}, except
for the exponential distribution, which is excluded for reasons given
in Sec.~\ref{sec_series}.  After introducing additional notations and
the random distribution functions in Sec.~\ref{sec:notation}, we give
a detailed explanation of the series generation in
Secs.~\ref{sec:graphexp} and \ref{sec:calc}, which should allow the
interested reader to follow each step.  As an example, we actually
show the complete calculation of a fourth order series in
Sec.~\ref{sec:example}.  In Sec.~\ref{sec_series} we present our
general-dimension series in full, accompanied by some discussion of
accuracy checks. Our series analysis and final results are described
in Sec.~\ref{sec:analysis} and we finish with our conclusions in
Sec.~\ref{sec:conclusion}.

\section{Further Notations and  Definitions}\label{sec:notation}

With $\beta=\frac{1}{\kb T}$, where $\kb$ denotes Boltzmann's constant
and $T$ the absolute temperature, the ensemble average of an
observable $A$ is calculated by
\begin{equation}\label{eq_thermav}
\langle A \rangle_T
=
\frac{
\Tr\left(A e^{-\beta\mathcal{H}} \right)
}{
Z
}
=
\frac{
\Tr\left(A e^{-\beta\mathcal{H}} \right)
}{
\Tr\left(e^{-\beta\mathcal{H}} \right)
}
\,,
\end{equation}
where the partition function $Z$ appears in the denominator.
Here the trace (Tr) is a shorthand for summing over all
possible values of the spins' $\{s_i\}$ configuration
\begin{equation}\label{eq_trace}
\Tr X=\Tr_{\{s_i\}} X(\{s_i\})=
\sum_{s_1=\pm1} \cdots
\sum_{s_N=\pm1}
X(\{s_i\})\,.
\end{equation}
The free energy per site $F$ is obtained from $Z$ by
\begin{equation}
F=\frac{1}{N}F_N\equiv-\frac{1}{\beta N}[\ln Z]_\mathrm{R}.
\end{equation}

Since the interaction constants $J_{ij}$ appear only in products with
$\beta$, it is convenient to use $\ka_{ij}=\beta J_{ij}$ as the
argument of the distribution functions introduced below. If $\Jcon^2$ is
some measure of $[J^2_{ij}]_\mathrm{R}$, then we also use $\ka=\beta
\Jcon$ as expansion variable, at least temporarily.  Since only even
powers of $\ka$ remain, we eventually use $\kex=\ka^2$ as the
expansion variable in our high temperature series.  Likewise we use
$x_\mathrm{c}=(\Jcon/(k_\mathrm{B}T_\mathrm{c}))^2$ to denote the {\em
  critical threshold}.

In the general case of a continuous probability distribution
$P(x)$, the configurational average is the nested integral
\begin{equation}
  [X]_\mathrm{R}=
  \idotsint\limits_{-\infty}^{-\infty}
  X(\{x_{ ij}\})
  \prod_{\langle ij \rangle } (P(x_{ ij})\multsp dx_{ ij})\,.
\end{equation}
For the bimodal random distribution the coupling constants
$\ka_{ij}$ for nearest neighbor pairs randomly assume only values of
either $+\ka$ or $-\ka$, so the latter integral can be written as the
nested sum
\begin{equation}
  [X]_\mathrm{R}= %
    \frac{1}{2^{Nd}}\sum\limits_{\{\ka_{\langle ij \rangle}=\pm \ka\}}
  X(\{\ka_{ij}\})\,,
\end{equation}
where a normalization factor of $1/2$ stems from each $\ka_{ij}$ in
the sum. In the $d$-dimensional hyper-cubic lattice with $N$ sites the
number of nearest neighbor pairs approaches $Nd$ for large $N$ when
boundary effects become negligible.

Near the critical temperature $T_\mathrm{c}$, the quantity of our
interest, $\EA$, is expected to
have a singularity of the form
\begin{equation}
  \label{eq:scaling-form}
  \EA\approx A(\pq - x)^{-\gamma}(1+B(\pq - x)^{\Delta_1}+\cdots).
\end{equation}
The aim of our analysis is to determine the critical exponent
$\gamma$ and, to a lesser extent, the first correction exponent
$\Delta_1$.  As for the free energy, we study this susceptibility per
lattice site.

\renewcommand{\timestamp}{Time-stamp: "2004-08-07 17:25:29 daboul"}
\subsection{The Random Distributions}\label{sec_random}

\begin{figure}
  \begin{center}
    \caption{The distribution functions for $\ka=5$.}\label{fig_distr}
      \includegraphics[width=.6\textwidth]{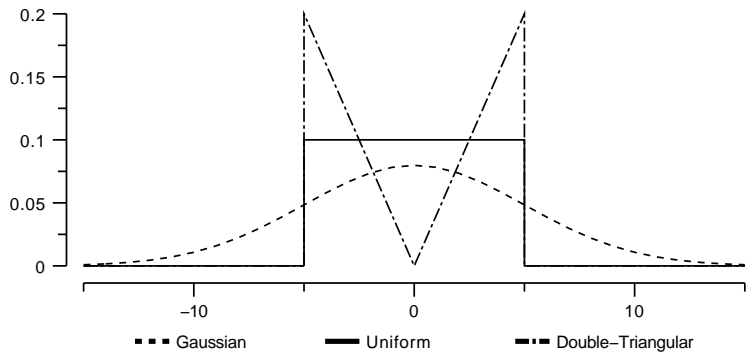}
  \end{center}
\end{figure}

\sts The different probability distributions, that we study,
are listed below.  We call them Bimodal, Gaussian,
Uniform and Double-Triangular.

\begin{eqnarray}
P_\mathrm{bim}(\rvar)&=& \frac{1}{2}(\delta[\rvar-\ka]+\delta[\rvar+\ka])
\\
P_\mathrm{gau}(\rvar)&=& \frac{{{e}^{-\frac{{\rvar^2}}{2 {\ka^2}}}}}{\ka
  {\sqrt{2 \pi }}}
\\
P_\mathrm{uni}(\rvar)&=& \left\{
\begin{array}{cc}
  1/(2 \ka) & \mbox{for } |\rvar|<\ka\\
  0 & \mbox{for }  |\rvar|\ge\ka
\end{array}\right.
\\
P_\mathrm{tri}(\rvar)&=& \left\{
\begin{array}{cc}
  |\rvar|/\ka^2 & \mbox{for }  |\rvar|<\ka\\
  0 & \mbox{for }  |\rvar|\ge\ka
\end{array} \right.
\end{eqnarray}
\begin{itemize}

\item The distributions are largely characterized by their moments
\be\label{eq_moment}
M_n\equiv[\rvar^n]_\mathrm{R} =\int_{-\infty}^\infty \rvar^n\ P(\rvar)\ d\rvar.
\ee

\item Since all distributions have the symmetry $P(-\rvar)=P(\rvar)$, the
  moments for odd $n$ vanish. In particular, the distributions have
  zero mean $[\rvar]_\mathrm{R}=\int_{-\infty}^\infty \rvar\ P(\rvar)\ d\rvar=0$.

\item For even $n$ the moments are:
  \begin{description}
  \item[Bimodal distribution:] $\ka^n$
  \item[Gaussian distribution:] $(n-1)!!\,\ka^n$
  \item[Uniform distribution:] $\ka^n/(n+1)$
  \item[Double-Triangular distribution:] $\ka^n/(n/2+1)$
  \end{description}

\item Thus all the distributions are normalized $\int_{-\infty}^\infty
  P(\rvar)\ d\rvar=1$.

\item A distribution's second moment $M_2 =\int_{-\infty}^\infty
  \rvar^2\ P(\rvar)\ d\rvar$, equal to the variance, is commonly
  associated with its width. In all cases it is linear in $\ka^2$, but
  with different pre-factors. Explicitly, $M_2$ is equal to $\ka^2$
  (Bimodal), $\ka^2$ (Gaussian), $\ka^2/3$ (Uniform), and $\ka^2/2$
  (Double-Triangular), respectively. With slightly redefined variables,
  $M_2$ could be equal to $\ka^2$ in all cases, which, in
  retrospective, would have been nicer.

\item Figure \ref{fig_distr} illustrates the distribution functions.
  The plots were calculated for the parameter $\ka=5$.

\end{itemize}

\subsection{Tangential Moments}\label{sec:moments}
In the calculation of the series we encounter the following integrals
over the distributions
\begin{equation}\label{eq:tangential}
  m_n\equiv[k^n]_\mathrm{R}=\int \tanh^n(\rvar)\ P(\rvar)\ d\rvar.
\end{equation}
We sometimes refer to $m_n$ as the {\em $n$-th tangential moment} of
the distribution, in order to distinguish it from the regular moment
(\ref{eq_moment}).  For a series to power $\ka^{2N}$ we need all
moments up to $m_N$ (not $2N$ as we will see later).
For the bimodal distribution the tangential moments are trivial:
$$
m_n=\frac{1}{2}(\tanh^n (\ka)+\tanh^n (-\ka))=
\left\{
\begin{array}{lr}
  0 & \mbox{ odd } n\\
  \tanh^n \ka & \mbox{ even }  n.
\end{array}
\right.
$$
The simple form $m_{2n}=\tanh^{2n}\ka=w^n$ makes $w\equiv\tanh^2(\ka)$ an alternative
(and convenient) expansion variable for this case, which has been used
in the past \cite{SinghC86,KleinAAHM91}. This, however, is not true
for the other distributions.  For them it may be possible to calculate
the $m_n$ analytically, as well. But the results may be complicated
functions of the parameter $\ka$, not suitable for our power series
expansion. Hence we are content to calculate the necessary moments
$m_n$ of all distributions as series in $\ka$. To avoid the tedious
work, this can conveniently be done with software for symbolic
computation, such as Mathematica. The obtained coefficients are later
used in the graph expansion, both for the computerized calculation and for
the example in this article. The symmetry of the distributions
makes all moments for odd $n$ vanish. Due to the power series of
$\tanh^n \ka$, each moment $m_n$ has only powers of $\ka^n$ and
higher, somewhat important during cumulant subtraction.  Note that in
the framework of series expansions, this is an exact treatment of the
randomness. We do not lose any additional information since a priori
we are limited to the highest order of our final series.

As an illustration, we show the expansion of the first few moments for
the bimodal and for the Gaussian distributions, to be used in the
example below.

\subsubsection{Bimodal Tangential Moments}
\begin{eqnarray}
m_0&=&1
\ ,\\
m_2&=&
{\ka^2}-\frac{2\multsp {\ka^4}}{3}+\frac{17\multsp {\ka^6}}{45}-\frac{62\multsp {\ka^8}}{315}+\frac{1382\multsp {\ka^{10}}}{14175}-\frac{21844\multsp
{\ka^{12}}}{467775}+  \nonumber\\
&& \frac{929569\multsp {\ka^{14}}}{42567525}-\frac{6404582\multsp {\ka^{16}}}{638512875}+\frac{443861162\multsp {\ka^{18}}}{97692469875}-\frac{18888466084\multsp
{\ka^{20}}}{9280784638125}+  \nonumber\\
&& \frac{113927491862\multsp {\ka^{22}}}{126109485376875}-\frac{58870668456604\multsp {\ka^{24}}}{147926426347074375}+\frac{8374643517010684\multsp
{\ka^{26}}}{48076088562799171875}-  \nonumber\\
&& \frac{689005380505609448\multsp
  {\ka^{28}}}{9086380738369043484375}+\frac{129848163681107301953\multsp {\ka^{30}}}{3952575621190533915703125}
\ ,\\
m_4&=&
{\ka^4}-\frac{4\multsp {\ka^6}}{3}+\frac{6\multsp {\ka^8}}{5}-\frac{848\multsp {\ka^{10}}}{945}+\frac{8507\multsp {\ka^{12}}}{14175}-  \nonumber\\
&& \frac{3868\multsp {\ka^{14}}}{10395}+\frac{46471426\multsp {\ka^{16}}}{212837625}-\frac{47060768\multsp {\ka^{18}}}{383107725}+\frac{518299498\multsp
{\ka^{20}}}{7753370625}-  \nonumber\\
&& \frac{92014385608\multsp {\ka^{22}}}{2598619698675}+\frac{39319617599924\multsp {\ka^{24}}}{2143861251406875}-\frac{12160377940064\multsp
{\ka^{26}}}{1304465840803125}+  \nonumber\\
&& \frac{14121349128787207129\multsp {\ka^{28}}}{3028793579456347828125}-\frac{20894145609681223868\multsp {\ka^{30}}}{9086380738369043484375}
\ ,\\
m_6&=&
{\ka^6}-2\multsp {\ka^8}+\frac{37\multsp {\ka^{10}}}{15}-\frac{2266\multsp {\ka^{12}}}{945}+\frac{1901\multsp {\ka^{14}}}{945}-\frac{79214\multsp {\ka^{16}}}{51975}+\frac{136750052\multsp
{\ka^{18}}}{127702575}-  \nonumber\\
&& \frac{64742312\multsp {\ka^{20}}}{91216125}+\frac{3282022\multsp {\ka^{22}}}{7309575}-\frac{710423622556\multsp {\ka^{24}}}{2598619698675}+\frac{82292419438259\multsp
{\ka^{26}}}{510443155096875}-  \nonumber\\
&& \frac{68433004067940682\multsp {\ka^{28}}}{739632131735371875}+\frac{157107220075270779857\multsp {\ka^{30}}}{3028793579456347828125}
\ .\end{eqnarray}

\subsubsection{Gaussian Tangential Moments}
\begin{eqnarray}
m_0&=&1
\ ,\\
m_2&=&
{\ka^2}-2\multsp {\ka^4}+\frac{17\multsp {\ka^6}}{3}-\frac{62\multsp {\ka^8}}{3}+\frac{1382\multsp {\ka^{10}}}{15}-\frac{21844\multsp {\ka^{12}}}{45}+\nonumber\\
&&\frac{929569\multsp {\ka^{14}}}{315}-\frac{6404582\multsp {\ka^{16}}}{315}+\frac{443861162\multsp {\ka^{18}}}{2835}-\frac{18888466084\multsp
{\ka^{20}}}{14175}+  \nonumber\\
&& \frac{1936767361654\multsp {\ka^{22}}}{155925}-\frac{58870668456604\multsp {\ka^{24}}}{467775}+\frac{8374643517010684\multsp {\ka^{26}}}{6081075}-
 \nonumber\\
&& \frac{689005380505609448\multsp {\ka^{28}}}{42567525}+\frac{129848163681107301953\multsp {\ka^{30}}}{638512875}
\ ,\\ %
m_4&=&
3\multsp {\ka^4}-20\multsp {\ka^6}+126\multsp {\ka^8}-848\multsp {\ka^{10}}+\frac{93577\multsp {\ka^{12}}}{15}-  \nonumber\\
&& 50284\multsp {\ka^{14}}+\frac{46471426\multsp {\ka^{16}}}{105}-\frac{800033056\multsp {\ka^{18}}}{189}+\frac{9847690462\multsp {\ka^{20}}}{225}-
 \nonumber\\
&& \frac{92014385608\multsp {\ka^{22}}}{189}+\frac{904351204798252\multsp {\ka^{24}}}{155925}-\frac{12160377940064\multsp {\ka^{26}}}{165}+
 \nonumber\\
&& \frac{14121349128787207129\multsp
  {\ka^{28}}}{14189175}-\frac{605930222680755492172\multsp
  {\ka^{30}}}{42567525}
\ ,\\
m_6&=&
15\multsp {\ka^6}-210\multsp {\ka^8}+2331\multsp {\ka^{10}}-24926\multsp {\ka^{12}}+271843\multsp {\ka^{14}}-  \nonumber\\
&& 3089346\multsp {\ka^{16}}+\frac{2324750884\multsp {\ka^{18}}}{63}-\frac{20911766776\multsp {\ka^{20}}}{45}+  \nonumber\\
&& 6173483382\multsp {\ka^{22}}-\frac{16339743318788\multsp {\ka^{24}}}{189}+\frac{1892725647079957\multsp {\ka^{26}}}{1485}-  \nonumber\\
&& \frac{68433004067940682\multsp {\ka^{28}}}{3465}+\frac{4556109382182852615853\multsp {\ka^{30}}}{14189175}\ .
\end{eqnarray}

\renewcommand{\timestamp}{Time-stamp: "2004-08-07 20:06:36 daboul"}
\section{Connected Graph Expansion and Cumulant Subtraction}
\label{sec:graphexp}

\sts An extensive physical quantity $X$\inthesis{ (having the {\em
    extensive property})} can be expanded in terms of connected graphs
only~\cite{SykesEHH66}. To order $n$ in a suitable expansion variable,
say $x$, all connected graphs with $n$ or less edges are used,
\begin{equation}\label{eq_graphexp}
  X=\sum_{b=0}^n
  \sum_{\Gamma\in\Gamma_b} w(\Gamma)\, X_\Gamma^\mathrm{c} +O(|x|^{n+1}).
\end{equation}
Here $\Gamma_b$ denotes the set of all connected graphs $\Gamma$ with
$b$ edges, $w(\Gamma)$ is the lattice constant for weak embeddings of
$\Gamma$ inside the lattice (see Sec.~\ref{sec_weakdef}) and
$X_\Gamma^\mathrm{c}$ is the cumulant of the graph's contribution.
For the quantity we calculate, the empty graph and the single vertex
graph (with 0 edges) together only contribute a constant summand of 1
with the chosen normalization.

The cumulant contribution of a connected graph $\Gamma$ is obtained by
subtracting off the cumulant contribution of all its connected
subgraphs,
\begin{equation}\label{eq_cumulant}
  X_\Gamma^\mathrm{c}= X_\Gamma - \sum_{\gamma\subset\Gamma} X_\gamma^c.
\end{equation}
Due to the subtractions, $X_\Gamma^\mathrm{c}$ is the contribution to
$X_\Gamma$, which depends on every one of the $b$ edges in $\Gamma$
and thus has only terms of order $b$ and higher in the expansion
variable.  This property allows us to stop the expansion at a certain
size of graphs, with a series which is correct to the obtained order,
and is in contrast to the original $X_\Gamma$, which can contribute to
any power.

We often use the term {\em bond} instead of {\em edge} and likewise
{\em site} instead of {\em vertex} since we deal with a physical model
on a lattice, and will eventually embed the graph inside it.  The
physical model is also the reason we do not address digraphs or graphs
with loops. Here the term {\em loop}, as commonly used in
graph-theory, denotes an edge whose both ends are incident on the
same vertex. This must be distinguished from a {\em cycle} (closed
path) in the graph, which {\em is} important to us.

Our model involves only nearest-neighbor interactions, visualized by
occupied lattice bonds. Since a graph $\Gamma$ on the lattice is
completely isolated from the rest of the infinite lattice by
unoccupied bonds, the thermodynamics of its spins is determined by the
reduced $N_\Gamma$-particle Hamiltonian for the graph
\begin{equation}
\mathcal{H}_{\Gamma}\{s_i\in\Gamma\}=-\sum_{\langle ij \rangle\in\Gamma} J_{ij}
s_i s_j.
\end{equation}

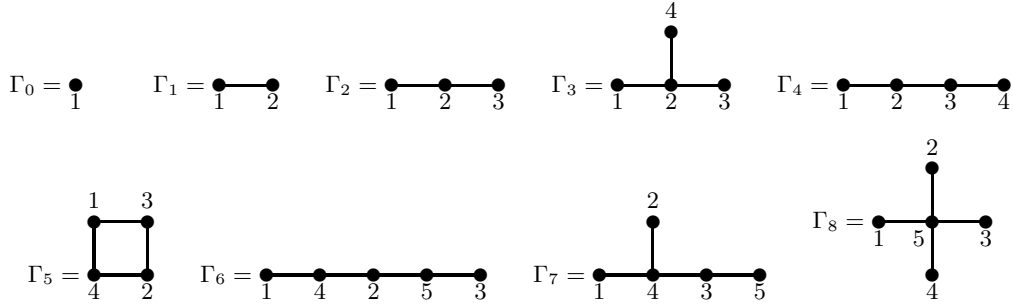
\begin{figure}
\caption{The smallest graphs, that can be embedded into the
hyper-cubic lattice.}\label{fig_graphs}
\begin{center}
\setlength{\unitlength}{0.007in}%

\begin{picture}(100,40)(190,544)
\thicklines\footnotesize
\put(210,555){\makebox(0,0)[cb]{\smash{$\Gamma_0=$}}}
\put(240,560){\circle*{10}}
\put(240,542){\makebox(0,0)[cb]{\smash{$1$}}}
\end{picture}
\begin{picture}(100,40)(190,544)
\thicklines\footnotesize
\put(210,555){\makebox(0,0)[cb]{\smash{$\Gamma_1=$}}}
\put(240,560){\circle*{10}}
\put(280,560){\circle*{10}}
\put(240,542){\makebox(0,0)[cb]{\smash{$1$}}}
\put(280,542){\makebox(0,0)[cb]{\smash{$2$}}}
\put(240,560){\line( 1, 0){40}}
\end{picture}
\quad
\begin{picture}(140,40)(190,544)
\thicklines\footnotesize
\put(210,555){\makebox(0,0)[cb]{\smash{$\Gamma_2=$}}}
\put(240,560){\circle*{10}}
\put(280,560){\circle*{10}}
\put(320,560){\circle*{10}}
\put(240,542){\makebox(0,0)[cb]{\smash{$1$}}}
\put(280,542){\makebox(0,0)[cb]{\smash{$2$}}}
\put(320,542){\makebox(0,0)[cb]{\smash{$3$}}}
\put(240,560){\line( 1, 0){80}}
\end{picture}
\quad
\begin{picture}(140,100)(190,544)
\thicklines\footnotesize
\put(210,555){\makebox(0,0)[cb]{\smash{$\Gamma_3=$}}}
\put(240,560){\circle*{10}}
\put(280,560){\circle*{10}}
\put(280,600){\circle*{10}}
\put(320,560){\circle*{10}}
\put(240,542){\makebox(0,0)[cb]{\smash{$1$}}}
\put(280,542){\makebox(0,0)[cb]{\smash{$2$}}}
\put(320,542){\makebox(0,0)[cb]{\smash{$3$}}}
\put(280,610){\makebox(0,0)[cb]{\smash{$4$}}}
\put(240,560){\line( 1, 0){80}}
\put(280,560){\line( 0, 1){40}}
\end{picture}
\quad
\begin{picture}(180,40)(190,544)
\thicklines\footnotesize
\put(210,555){\makebox(0,0)[cb]{\smash{$\Gamma_4=$}}}
\put(240,560){\circle*{10}}
\put(280,560){\circle*{10}}
\put(320,560){\circle*{10}}
\put(360,560){\circle*{10}}
\put(240,542){\makebox(0,0)[cb]{\smash{$1$}}}
\put(280,542){\makebox(0,0)[cb]{\smash{$2$}}}
\put(320,542){\makebox(0,0)[cb]{\smash{$3$}}}
\put(360,542){\makebox(0,0)[cb]{\smash{$4$}}}
\put(240,560){\line( 1, 0){120}}
\end{picture}

\noindent\begin{picture}(100,100)(190,544)
\thicklines\footnotesize
\put(210,555){\makebox(0,0)[cb]{\smash{$\Gamma_5=$}}}
\put(240,560){\circle*{10}}
\put(280,560){\circle*{10}}
\put(280,600){\circle*{10}}
\put(240,600){\circle*{10}}
\put(240,542){\makebox(0,0)[cb]{\smash{$4$}}}
\put(280,542){\makebox(0,0)[cb]{\smash{$2$}}}
\put(240,610){\makebox(0,0)[cb]{\smash{$1$}}}
\put(280,610){\makebox(0,0)[cb]{\smash{$3$}}}
\put(240,560){\line( 1, 0){40}}
\put(240,560){\line( 0, 1){40}}
\put(240,600){\line( 1, 0){40}}
\put(280,560){\line( 0, 1){40}}
\end{picture}
\quad
\noindent\begin{picture}(220,40)(190,544)
\thicklines\footnotesize
\put(210,555){\makebox(0,0)[cb]{\smash{$\Gamma_6=$}}}
\put(240,560){\circle*{10}}
\put(280,560){\circle*{10}}
\put(320,560){\circle*{10}}
\put(360,560){\circle*{10}}
\put(400,560){\circle*{10}}
\put(240,542){\makebox(0,0)[cb]{\smash{$1$}}}
\put(280,542){\makebox(0,0)[cb]{\smash{$4$}}}
\put(320,542){\makebox(0,0)[cb]{\smash{$2$}}}
\put(360,542){\makebox(0,0)[cb]{\smash{$5$}}}
\put(400,542){\makebox(0,0)[cb]{\smash{$3$}}}
\put(240,560){\line( 1, 0){160}}
\end{picture}
\quad
\begin{picture}(180,100)(190,544)
\thicklines\footnotesize
\put(210,555){\makebox(0,0)[cb]{\smash{$\Gamma_7=$}}}
\put(240,560){\circle*{10}}
\put(280,560){\circle*{10}}
\put(280,600){\circle*{10}}
\put(320,560){\circle*{10}}
\put(360,560){\circle*{10}}
\put(240,542){\makebox(0,0)[cb]{\smash{$1$}}}
\put(280,542){\makebox(0,0)[cb]{\smash{$4$}}}
\put(320,542){\makebox(0,0)[cb]{\smash{$3$}}}
\put(360,542){\makebox(0,0)[cb]{\smash{$5$}}}
\put(280,610){\makebox(0,0)[cb]{\smash{$2$}}}
\put(240,560){\line( 1, 0){120}}
\put(280,560){\line( 0, 1){40}}
\end{picture}
\quad
\begin{picture}(140,140)(190,504)
\thicklines\footnotesize
\put(210,555){\makebox(0,0)[cb]{\smash{$\Gamma_8=$}}}
\put(240,560){\circle*{10}}
\put(280,560){\circle*{10}}
\put(280,600){\circle*{10}}
\put(280,520){\circle*{10}}
\put(320,560){\circle*{10}}
\put(240,542){\makebox(0,0)[cb]{\smash{$1$}}}
\put(270,542){\makebox(0,0)[cb]{\smash{$5$}}}
\put(280,502){\makebox(0,0)[cb]{\smash{$4$}}}
\put(320,542){\makebox(0,0)[cb]{\smash{$3$}}}
\put(280,610){\makebox(0,0)[cb]{\smash{$2$}}}
\put(240,560){\line( 1, 0){80}}
\put(280,560){\line( 0, 1){40}}
\put(280,560){\line( 0,-1){40}}
\end{picture}

\end{center}
\end{figure}

\subsection{Cumulant Subtraction}\label{sec_cumdef}

Equation (\ref{eq_cumulant}) contains the sum over connected
subgraphs.  In the following we have written out the cumulants for the
smallest graphs, with explicit numerical coefficients and graph
indices, because we will need them in the example later.  For these
small graphs, the expressions can easily be confirmed by visual
inspection using Fig.~\ref{fig_graphs}. These cumulants are given by
\newcommand{\C}[1]{X^\mathrm{c}_{\Gamma_{#1}}}
\newcommand{\U}[1]{X_{\Gamma_{#1}}}
\begin{equation}
\begin{array}{rcl}
  \C{0}&=&\U{0},\\
  \C{1}&=&\U{1} - 2\C{0},\\
  \C{2}&=&\U{2} - 3\C{0}- 2\C{1},\\
  \C{3}&=&\U{3} - 4\C{0}- 3\C{1}- 3\C{2},\\
  \C{4}&=&\U{4} - 4\C{0}- 3\C{1}- 2\C{2},\\
  \C{5}&=&\U{5} - 4\C{0}- 4\C{1}- 4\C{2}- 4\C{4},\\
  \C{6}&=&\U{6} - 5\C{0}- 4\C{1}- 3\C{2}- 2\C{4},\\
  \C{7}&=&\U{7} - 5\C{0}- 4\C{1}- 4\C{2}- 1\C{3}- 2\C{4},\\
  \C{8}&=&\U{8} - 5\C{0}- 4\C{1}- 6\C{2}- 4\C{3}.
\end{array}
\end{equation}

\subsection{The Lattice Constants for Weak Embeddings}\label{sec_weakdef}

In (\ref{eq_graphexp}), the cumulant contribution of each graph is
multiplied by its {\em lattice constant} $w(\Gamma)$. This constant is
the number of distinct ways per lattice-site in which the graph can be
weakly embedded in a particular lattice, and thus it ties our series
to that specific lattice. In this \inthesis{work}{article} we address
the $d$-dimensional hyper-cubic lattice, which, as the term suggests,
is a generalization of the square lattice ($d=2$) and the cubic
lattice ($d=3$). We use the tabulated functions $w_\Gamma(d)$
from~\cite{wan_91}.  To calculate our example later on, we need the
lattice constants of the first graphs:
\begin{equation}
\begin{array}{rcl}
  w(\Gamma_0)&=&
  1,\\
  w(\Gamma_1)&=&
  d,\\
  w(\Gamma_2)&=&
  -d+2\multsp {d^2},\\
  w(\Gamma_3)&=&
  \frac{2\multsp d}{3}-2\multsp {d^2}+\frac{4\multsp {d^3}}{3},\\
  w(\Gamma_4)&=&
  d-4\multsp {d^2}+4\multsp {d^3},\\
  w(\Gamma_5)&=&
  -\frac{d}{2}+\frac{{d^2}}{2},\\
  w(\Gamma_6)&=&
  d+4\multsp {d^2}-12\multsp {d^3}+8\multsp {d^4},\\
  w(\Gamma_7)&=&
  -2\multsp d+10\multsp {d^2}-16\multsp {d^3}+8\multsp {d^4},\\
  w(\Gamma_8)&=&
  -\frac{d}{2}+\frac{11\multsp {d^2}}{6}-2\multsp {d^3}+
  \frac{2\multsp {d^4}}{3}.
\end{array}
\end{equation}

\section{General Calculation and Simplifications for the Ising Spin
  Glass\label{sec:calc}}

The Boltzmann factor $e^{-\beta\mathcal{H}}$ can be rewritten in a way
that makes the calculation of the trace more convenient. With the notations
$\lambda=\beta h_0$, $\tau=\tanh\lambda$, $K_{ij}=\cosh \ka_{ij}$ and $\kc_{ij}=\tanh \ka_{ij}$ we obtain:
\begin{eqnarray}
  e^{-\beta\mathcal{H}}
  &=&
  \exp\left({\beta\sum_{\langle ij \rangle} J_{ij}
      s_i s_j \ +
      \beta h_0\sum_{i=1}^N  s_i}\right)
  \nonumber \\
  \inthesis{
    &=&
    \prod_{\langle ij \rangle } \exp(\ka_{ij} s_i s_j )
    \prod_{i=1}^N \exp(\lambda s_i)
    \nonumber\\}{}
  &=&
  \prod_{\langle ij \rangle }
  \left( \cosh(\ka_{ij} s_i s_j )+\sinh(\ka_{ij} s_i s_j ) \right)
  \prod_{i=1}^N
  \left( \cosh(\lambda s_i)+\sinh(\lambda s_i) \right).
\end{eqnarray}
Now we exploit the fact that the only possible values of $s_i$ are $\pm1$,
together with the symmetry of $\cosh$ and anti-symmetry of $\sinh$:
\inthesis{
  \begin{eqnarray}
    e^{-\beta\mathcal{H}}
    &=&
    \prod_{\langle ij \rangle } \left( \cosh \ka_{ij}  \left( 1+s_i s_j
        \tanh \ka_{ij} \right)\right)
    \ \cosh^N\lambda
    \ \prod_{i=1}^N
    \left(1+s_i\tanh\lambda \right)
    \\
    &=&
    \cosh^N\lambda\
    \prod_{\langle ij \rangle } K_{ij}
    \prod_{\langle ij \rangle }  \left( 1+s_i s_j \kc_{ij} \right)
    \ \prod_{i=1}^N \left(1+s_i\tau \right).
  \end{eqnarray}}{
  \begin{equation}
    e^{-\beta\mathcal{H}}
    =
    \cosh^N\lambda\
    \prod_{\langle ij \rangle } K_{ij}
    \prod_{\langle ij \rangle }  \left( 1+s_i s_j \kc_{ij} \right)
    \ \prod_{i=1}^N \left(1+s_i\tau \right).
  \end{equation}}
\inthesis{We address the case without external magnetic field,
  $h_0=0$, where the Boltzmann factor simplifies to
  \begin{equation}
    e^{-\beta\mathcal{H}}
    =
    \prod_{\langle ij \rangle } K_{ij}
    \prod_{\langle ij \rangle }  \left( 1+s_i s_j \kc_{ij} \right).
  \end{equation}
  We now use the derived equation for the numerator and denominator in
  the calculation of an observable $\langle A \rangle_T$
  (\ref{eq_thermav}):
\begin{equation}
\Tr\left(A e^{-\beta\mathcal{H}} \right)
=
\cosh^N\lambda\
\prod_{\langle ij \rangle } K_{ij}\ 2^N
\underbrace{\frac{1}{2^N} \Tr\left(A\prod_{\langle ij \rangle }  \left( 1+\kc_{ij}s_i s_j
\right)\ \prod_{i=1}^N \left(1+s_i\tau \right)\right)}_{X_\mathrm{R}}
\end{equation}
\begin{equation}
Z=\Tr\left(e^{-\beta\mathcal{H}} \right)
=
\cosh^N\lambda\
\prod_{\langle ij \rangle } K_{ij}\ 2^N
\underbrace{\frac{1}{2^N} \Tr\left(\prod_{\langle ij \rangle }  \left( 1+\kc_{ij}s_i s_j
\right)\ \prod_{i=1}^N \left(1+s_i\tau \right)\right)}_{Z_\mathrm{R}}
\end{equation}
We call}{%
We address the case without external magnetic field,
$h_0=0$, and call}
\begin{equation}\label{eq_zr}
Z_\mathrm{R}=\frac{1}{2^N} \Tr\left(\prod_{\langle ij \rangle }  \left( 1+\kc_{ij}s_i s_j
\right)\ \prod_{i=1}^N \left(1+s_i\tau \right)\right)
\end{equation}
the {\em reduced partition function} and sometimes use the notation
$Z_\mathrm{R}(\Gamma_n)=Z_{\Gamma_n}$.  For the case of zero external
magnetic field $h_0=0$ we end up with the important (for our coming
calculations) equation
\begin{equation}\label{eq:observable}
\langle A \rangle_T=
\frac{
  \Tr_{\{s_i\}}\left(A\prod_{\langle ij \rangle }
  \left( 1+\kc_{ij}s_i s_j\right)\right)
}{
  \Tr_{\{s_i\}}\left(\prod_{\langle ij \rangle }
  \left( 1+\kc_{ij}s_i s_j \right)\right)
}.
\end{equation}
Often $A$ will be a linear combination of $s_m s_n$. We will then see
the terms
\begin{equation}
\langle s_m s_n \rangle_T= \frac{ 2^{-N} \Tr_{\{s_i\}}\left(s_m
  s_n\prod_{\langle ij \rangle } \left( 1+\kc_{ij}s_i
  s_j\right)\right) }{ 2^{-N} \Tr_{\{s_i\}}\left(\prod_{\langle ij
    \rangle } \left( 1+\kc_{ij}s_i s_j\right)\right) }.
\end{equation}

\subsection{Graph-Expansion for the Edwards-Anderson Susceptibility}

For the graph expansion of the Edwards-Anderson susceptibility
(\ref{eq:EA}) we apply the general formula (\ref{eq_graphexp}).  The
calculation is done to order $n$ in $\kex=\ka^2$. With this expansion
variable all the dependence on the coupling strength parameter $\Jcon$
and the temperature $T$ are absorbed in the argument of the power
series. We obtain
\begin{equation}
N \EA= \sum_{b=0}^n \sum_{\Gamma\in\Gamma_b} w(\Gamma)
X_\Gamma^c.
\end{equation}
and also denote the associated (non-cumulant) observable on only one
(sub)graph as
\begin{equation}\label{eq_xgammasum}
X_\Gamma
= \sum_{i,j\in\Gamma} \left[\langle
s_i s_j \rangle^2_T\right]_\mathrm{R}.
\end{equation}

\section{Explicit Calculation for the Smallest Graphs}\label{sec:example}

We now show explicitly the calculation of $\EA$ for the smallest
graphs in Fig.~\ref{fig_graphs}, using the equations from the previous
section. Here we often use $ \left[ \kc_{ij}^n\right]_\mathrm{R}= m_n$
following definition (\ref{eq:tangential}).

\setlength{\unitlength}{0.007in}%

\begin{itemize}
\item For {$\Gamma_0$}, we have
\begin{equation}
\EA(\Gamma_0)=
1 \left[\langle s_0^2 \rangle^2_T\right]_\mathrm{R}=
\left[\langle 1 \rangle^2_T\right]_\mathrm{R}=1.
\end{equation}
For any spin, the self-correlation trivially equals 1.

\item For $\Gamma_1$,
  \begin{equation}
    \EA(\Gamma_1)= \left[\langle s_1 s_2
      \rangle^2_T\right]_\mathrm{R}+\left[\langle s_2 s_1
      \rangle^2_T\right]_\mathrm{R} + \langle s_1\rangle^2+ \langle
    s_2\rangle^2 =2 \left[\langle s_1 s_2
      \rangle^2_T\right]_\mathrm{R} +2.
  \end{equation}
  The denominator of $\langle s_1 s_2 \rangle_T$ is
  \begin{equation}
    Z_{\Gamma_1}=Z_R({\Gamma_1})=2^{-2}\Tr_{\{s_1,s_2\}}\left(1+\kc_{12}s_1 s_2\right)=1,
  \end{equation}
  and the numerator is
  \begin{equation}
    \langle s_1 s_2 \rangle_T\ Z_{\Gamma_1}=
    2^{-2} \Tr_{\{s_1,s_2\}}\left(s_1 s_2(1+\kc_{12}s_1 s_2)\right)=
    2^{-2} \Tr_{\{s_1,s_2\}}\left(s_1 s_2+\kc_{12}\right)=
    \kc_{12}.
  \end{equation}
  Thus,
  \begin{equation}
    \EA(\Gamma_1)=2 \left[\langle s_1 s_2 \rangle^2_T\right]_\mathrm{R} +2=
    2 \left[\kc_{12}^2\right]_\mathrm{R}+2
    =2 \left[\tanh^2(\beta J_{12})\right]_\mathrm{R}+2
    =2 m_2+2,
  \end{equation}
  where we used the definition of the tangential moment $m_2$.

\item For $\Gamma_2$,
\begin{equation}
\EA(\Gamma_2)=
2 \left(
 \left[\langle s_1 s_2 \rangle^2_T\right]_\mathrm{R} +
 \left[\langle s_2 s_3 \rangle^2_T\right]_\mathrm{R} +
 \left[\langle s_1 s_3 \rangle^2_T\right]_\mathrm{R}
\right)+3.
\end{equation}
The denominator of each correlation is
\begin{eqnarray}
Z_{\Gamma_2}&=&
2^{-3} \Tr_{\{s_1,s_2,s_3\}}\left(
(1+\kc_{12}s_1 s_2)
(1+\kc_{23}s_2 s_3)
\right)\\
&=&
2^{-3} \Tr_{\{s_1,s_2,s_3\}}\left(
1+\kc_{12}s_1 s_2+\kc_{23}s_2 s_3+\kc_{12}\kc_{23}s_1 s_3
\right)=1.
\end{eqnarray}
The numerator is
\begin{equation} \langle s_1 s_2 \rangle_T\ Z_{\Gamma_2}= 2^{-3}
\Tr_{\{s_1,s_2,s_3\}}\left(s_1 s_2(1+\kc_{12}s_1 s_2)(1+\kc_{23}s_2
s_3)\right)= \kc_{12},
\end{equation}
and likewise
\begin{equation}
\langle s_2 s_3 \rangle_T \
Z_{\Gamma_2} = \kc_{23}\, \mbox{, and}
\end{equation}
\begin{equation} \langle s_1 s_3 \rangle_T\ Z_{\Gamma_2}
= 2^{-3} \Tr_{\{s_1,s_2,s_3\}}\left(s_1 s_3(1+\kc_{12}s_1 s_2)
(1+\kc_{23}s_2 s_3)\right) = \kc_{12} \kc_{23}.
\end{equation}
The configurational average gives
\begin{equation}
\left[\langle s_1 s_2
  \rangle^2_T\right]_\mathrm{R} = \left[
  \kc_{12}^2\right]_\mathrm{R}=m_2\,, \quad \left[\langle s_2 s_3
  \rangle^2_T\right]_\mathrm{R} = m_2 \quad \mbox{and}
\end{equation}
\begin{equation}
\left[\langle s_1 s_3
  \rangle^2_T\right]_\mathrm{R} = \left[ \kc_{12}^2
  \kc_{23}^2\right]_\mathrm{R}= \left[ \kc_{12}^2\right]_\mathrm{R}
\left[ \kc_{23}^2\right]_\mathrm{R}= m_2^2\,,
\end{equation}
where in the last step we used the fact that the random variables
$\kc_{ij}$ for different bonds are uncorrelated. Finally,
\begin{equation}
\EA(\Gamma_2)=3+ 4 m_2+ 2 m_2^2.
\end{equation}

\end{itemize}
A few general observations are in order: From the possible values of a
spin variable, $s_i=\pm1$, we  trivially have $s_i^2=1$.  Further, the
trace sums over the possible values of each spin, so any
summand with an unpaired $s_i$ vanishes. When writing out the product
within the traces $(1+\kc_{ij}s_i s_j)\cdots(1+\kc_{kl}s_k s_l)$ one
has to choose from each pair of $\langle ij \rangle$ (and for any
resulting summand) either the constant 1 or the bond factor $\kc_{ij}$. To
find the terms that will actually survive the trace, the $s_i s_j$
site-factors accompanying any bond have to be combined with other
sites, either from another bond in the graph or from terms originally
present inside the trace.

For any acyclic graph, which we have seen so far, we immediately see
that the reduced partition function is always equal to 1; no bond can
be combined with others to eliminate all unpaired sites. Also the
numerator for pair correlations remains simple. Visually only a path
of bonds joining the two sites in question yields factors of only
paired spins, which survive the trace. Each such bond contributes a
factor of $m_2$. A constant multiplier results from the number of ways
the pair of sites (and equivalent pairs) can be joined.  For cyclic
graphs the calculation becomes much more complicated.
In passing please note that one of the inherent features of spin
glasses is {\em frustration}. Only a cyclic graph is susceptible to
this phenomenon and as such can possibly integrate true SG-properties
into our series.
\begin{itemize}
\item For $\Gamma_3$,
  \begin{eqnarray}
    \EA(\Gamma_3)&=&
2 \left[
\langle s_1 s_2 \rangle^2_T +
\langle s_2 s_3 \rangle^2_T +
\langle s_2 s_4 \rangle^2_T +
\langle s_1 s_3 \rangle^2_T +
\langle s_1 s_4 \rangle^2_T +
\langle s_3 s_4 \rangle^2_T \right]_\mathrm{R}
+4\nonumber\\
&=&
4+6m_2+6m_2^2.
\end{eqnarray}

\item For
{$\Gamma_4$},
\begin{eqnarray}
\EA(\Gamma_4)&=&
2 \left(\;
 \left[\langle s_1 s_2 \rangle^2_T\right]_\mathrm{R} +
 \left[\langle s_2 s_3 \rangle^2_T\right]_\mathrm{R} +
 \left[\langle s_3 s_4 \rangle^2_T\right]_\mathrm{R} +\right.\nonumber\\
& &\left.\phantom{2(\;}
 \left[\langle s_1 s_3 \rangle^2_T\right]_\mathrm{R} +
 \left[\langle s_2 s_4 \rangle^2_T\right]_\mathrm{R} +
 \left[\langle s_1 s_4 \rangle^2_T\right]_\mathrm{R}
 \;\right)+4\nonumber\\
&=&
4+6m_2+4m_2^2+2m_2^3.
\end{eqnarray}

\item For
{$\Gamma_5$},
\begin{equation}\label{eq:gam5}
\EA(\Gamma_5)=
4+
8 \left[\langle s_1 s_3 \rangle^2_T\right]_\mathrm{R} +
4 \left[\langle s_1 s_2 \rangle^2_T\right]_\mathrm{R}.
\end{equation}
Here we used the equivalence of {\em pairs of sites}, in terms of
adjacencies, to reduce the number of terms to 3. We defer calculation
of the result till later.

\item For
{$\Gamma_6$},
\begin{equation}
\EA(\Gamma_6)=
{2} \left[
4 \langle s_1 s_4 \rangle^2_T +
3 \langle s_1 s_2 \rangle^2_T +
2 \langle s_1 s_5 \rangle^2_T +
1 \langle s_1 s_3 \rangle^2_T
\right]_\mathrm{R} +5
=
5 + 8 m_2+ 6 m_2^2 + 4 m_2^3 + 2 m_2^4.
\end{equation}

\item For
{$\Gamma_7$},
\begin{equation}
\EA(\Gamma_7)=
2
\left[
4 \langle s_1 s_4 \rangle^2_T +
4 \langle s_1 s_2 \rangle^2_T +
2 \langle s_1 s_5 \rangle^2_T
\right]_\mathrm{R}
+5
=
5 + 8 m_2 + 8 m_2^2 + 4 m_2^3.
\end{equation}

\item For
{$\Gamma_8$},
\begin{equation}
\EA(\Gamma_8)=
2
\left[
4 \langle s_1 s_5 \rangle^2_T+
6 \langle s_1 s_2 \rangle^2_T
\right]_\mathrm{R}
+5
=
5 + 8 m_2 + 12 m_2^2.
\end{equation}

\end{itemize}

We now resume the calculation of $\EA(\Gamma_{5})$ from
(\ref{eq:gam5}) which, due to
the graph's cycle, is significantly more complicated than the
contribution of the other graphs. The complexity enters through the
non-trivial partition function
\begin{equation}
Z_{\Gamma_5}=
2^{-4} \Tr_{\{s_1,s_2,s_3,s_4\}}\left(
(1+\kc_{13}s_1 s_3)
(1+\kc_{23}s_2 s_3)
(1+\kc_{14}s_1 s_4)
(1+\kc_{24}s_2 s_4)
\right)=
1+\kc_{13}\kc_{23}\kc_{14}\kc_{24}.
\end{equation}
The numerators for the spin correlations are
$ \langle s_1 s_3
\rangle_T\ Z_{\Gamma_5}= \kc_{13}+\kc_{14}\kc_{24}\kc_{23}$ and $
\langle s_1 s_2 \rangle_T\ Z_{\Gamma_5}=
\kc_{13}\kc_{23}+\kc_{14}\kc_{24}$.
The complicated part is now performing the averaging over the randomness for
\begin{equation}
\left[\langle s_1 s_3 \rangle^2_T\right]_\mathrm{R}=
\left[\left(
\frac{
\kc_{13}+\kc_{14}\kc_{24}\kc_{23}
}{
1+\kc_{13}\kc_{23}\kc_{14}\kc_{24}
}
\right)^2\right]_\mathrm{R}
\end{equation}and
\begin{equation}\label{eq:ExplCalcG8a}
\left[\langle s_1 s_2 \rangle^2_T\right]_\mathrm{R}=
\left[\left(
\frac{
\kc_{13}\kc_{23}+\kc_{14}\kc_{24}
}{
1+\kc_{13}\kc_{23}\kc_{14}\kc_{24}
}
\right)^2\right]_\mathrm{R}
.
\end{equation}
In fact, for most probability distributions we do not know to
calculate this directly. For the bimodal distribution it is possible,
and was in fact done also for much larger graphs with a computer
\cite{KleinAAHM91}. For continuous distributions the calculation would
at best become extremely tedious and most likely not feasible for
large graphs. Our solution again utilizes a power expansion. By this
we do not lose any more information since our final series are limited
to a certain order in the expansion variable anyway.

To make the expansion process more obvious we rewrite the equations
with the symbols $\ku_{ij}={\kc_{ij}}/\kx$ such that $\kc_{ij}=\kx\, \ku_{ij}$ and
\begin{equation}
\langle s_1 s_3 \rangle_T=
\frac{
\kx\ku_{13}+\kx^3 \ku_{14}\ku_{24}\ku_{23}
}{
1+\kx^4 \ku_{13}\ku_{23}\ku_{14}\ku_{24}
},
\end{equation}
\begin{equation}
\langle s_1 s_2 \rangle_T=
\frac{
\kx^2 \ku_{13}\ku_{23}+ \kx^2 \ku_{14}\ku_{24}
}{
1+\kx^4 \ku_{13}\ku_{23}\ku_{14}\ku_{24}
},
\end{equation}
and expand the squares in powers of $\kx$:
\begin{eqnarray}
\langle s_1 s_3 \rangle^2_T&=& u_{13}^{2} {\kx^2}+2 {u_{13}} {u_{14}}
{u_{23}} {u_{24}} {\kx^4}+(-2 u_{13}^{3} {u_{14}} {u_{23}}
{u_{24}}+u_{14}^{2} u_{23}^{2} u_{24}^{2}) {\kx^6}\nonumber\\
 &&-4 (u_{13}^{2}
u_{14}^{2} u_{23}^{2} u_{24}^{2}) {\kx^8}+(3 u_{13}^{4} u_{14}^{2}
u_{23}^{2} u_{24}^{2}-2 {u_{13}} u_{14}^{3} u_{23}^{3} u_{24}^{3})
{\kx^{10}}\nonumber\\
&&+6 u_{13}^{3} u_{14}^{3} u_{23}^{3} u_{24}^{3}
{\kx^{12}}+(-4 u_{13}^{5} u_{14}^{3} u_{23}^{3} u_{24}^{3}+3 u_{13}^{2}
u_{14}^{4} u_{23}^{4} u_{24}^{4}) {\kx^{14}}+O(|\kx|^{15}),
\label{eq_correlation}
\end{eqnarray}
\begin{eqnarray}
\langle s_1 s_2 \rangle^2_T&=&
{\kx^4} (u_{13}^{2} u_{23}^{2}+2 {u_{13}} {u_{14}} {u_{23}} {u_{24}}+u_{14}^{2} u_{24}^{2})\nonumber\\
&&+ {\kx^8} (-2 u_{13}^{3} {u_{14}} u_{23}^{3} {u_{24}}-4 u_{13}^{2} u_{14}^{2} u_{23}^{2}
u_{24}^{2}-2 {u_{13}} u_{14}^{3} {u_{23}} u_{24}^{3})\nonumber\\
&&+ {\kx^{12}} (3 u_{13}^{4} u_{14}^{2} u_{23}^{4} u_{24}^{2}+6 u_{13}^{3} u_{14}^{3}
u_{23}^{3} u_{24}^{3}+3 u_{13}^{2} u_{14}^{4} u_{23}^{2} u_{24}^{4})+O(|\kx|^{15}).
\end{eqnarray}
With the fractions removed, we can again factorize for averages
over independent variables, and thus use the previously defined
moments of the random distributions:
\begin{eqnarray}
\left[\langle s_1 s_3 \rangle^2_T\right]_\mathrm{R}&=&
m_2+m_2^3-4m_2^4+3m_2^3m_4+3m_2m_4^3+O(|\ka|^{15}),\\
\left[\langle s_1 s_2 \rangle^2_T\right]_\mathrm{R}&=&
2m_2^2-4m_2^4+6m_2^2m_4^2+O(|\ka|^{15}).
\end{eqnarray}
Here we expanded to a higher power than actually necessary for the
largest graph that we consider in this example. It shows that higher
moments actually show up. We first encountered $m_3=0$ and
\begin{equation}
m_4=\left[\kc_{ij}^4\right]_\mathrm{R}
=\left[\tanh^4\ka_{ij}\right]_\mathrm{R}
=\int \tanh^4x\ P(x) dx.
\end{equation}
For the bimodal distribution we can quickly use the simple form of the
moments $m_2=w$ and $m_4=w^2$, and obtain
\begin{eqnarray}\label{eq:sisjsqcontrib}
\left[\langle s_1 s_3 \rangle^2_T\right]_\mathrm{R}&=&
w+w^3-4w^4+3w^5+3w^7+\ldots\\
\left[\langle s_1 s_2
\rangle^2_T\right]_\mathrm{R}&=& 2w^2-4w^4+6w^6+\ldots,
\end{eqnarray}
and thus
\begin{equation}
\EA(\Gamma_{5})=4+8w+8w^2+8w^3-48w^4+24w^5+24w^6+24w^7+O(|w|^8).
\end{equation}
For the general distribution we remain with
\begin{equation}
\EA(\Gamma_{5})=4+8(
m_2
+m_2^2
+m_2^3
-6m_2^4
+3m_2^3 m_4
+3m_2^2 m_4^2
+3m_2 m_4^3
)+O(|\ka|^{15}),
\end{equation}
on which we elaborate further.
\subsection{Performing Cumulant Subtraction}

In this section we perform the cumulant subtraction for the quantity
$\EA$, which we calculated in the previous section. The general
equations were given in Sec.~\ref{sec_cumdef} and in the following we
show the cumulant graph contributions first in terms of the tangential
moments and also for the case of the bimodal distribution using $w$ as
the expansion variable, thus substituting $m_2=w$ and $m_4=w^2$.
\begin{equation}
\begin{array}{rcl}
  \C{0}&=&1,\\
  \C{1}&=&2m_2=2w,\\
  \C{2}&=&2m_2^2=2w^2,\\
  \C{3}&=&0,\\
  \C{4}&=&2m_2^3=2w^3,\\
  \C{5}&=&-48m_2^4+24m_2^3m_4+24m_2^2m_4^2+24m_2m_4^3+\ldots\\
  &=&-48w^4+24w^5+24w^6+24w^7+\ldots,\\
  \C{6}&=&2m_2^4=2w^4,\\
  \C{7}&=&0,\\
  \C{8}&=&0.
\end{array}
\end{equation}
Indeed we see that no graph contributes to a power of $w$ less than
its number of bonds. This fact is used in the computerized calculation
as an internal check.

In general we use $\ka$ (or $x=\ka^2$) as the expansion variable, for
which we now use the expansions of the moments $m_n$ from
section \ref{sec:moments}. We show the result for the Gaussian
distribution:
\begin{equation}
\begin{array}{rcl}
  \C{0}&=&1,\\
  \C{1}&=&2\multsp {\ka^2}-4\multsp {\ka^4}+
  \frac{34\multsp {\ka^6}}{3}-\frac{124\multsp {\ka^8}}{3}+O(|\ka|^{10}),\\
  \C{2}&=&2\multsp {\ka^4}-8\multsp {\ka^6}+
  \frac{92\multsp {\ka^8}}{3}+O(|\ka|^{10}),\\
  \C{3}&=&0,\\
  \C{4}&=&2\multsp {\ka^6}-12\multsp {\ka^8}+O(|\ka|^{10}),\\
  \C{5}&=&-48\multsp {\ka^8}+O(|\ka|^{10}),\\
  \C{6}&=&2\multsp {\ka^8}+O(|\ka|^{10}),\\
  \C{7}&=&0,\\
  \C{8}&=&0.
\end{array}
\end{equation}

\subsection{Using the Lattice Constants}

Using (\ref{eq_graphexp}) together with the lattice constants
of the smallest graphs, as given in Sec.~\ref{sec_weakdef}, we can
now perform the final summation and obtain the series. For the
Gaussian distribution, we find
\begin{eqnarray}
\EA&=&
1+2\multsp d\multsp {\ka^2}+(-6\multsp d+4\multsp
{d^2})\multsp {\ka^4}+ \Big(\frac{64\multsp d}{3}-24\multsp
{d^2}+8\multsp {d^3}\Big)\multsp {\ka^6} \nonumber\\
&&+\bigg(-58\multsp
d+\frac{280\multsp {d^2}}{3}-72\multsp {d^3}+16\multsp
{d^4}\bigg)\multsp {\ka^8}+\ldots
\end{eqnarray}
For the bimodal distribution one has
\begin{eqnarray}
\EA&=&
1+2\multsp d\multsp {\ka^2}+\Big(-\frac{10\multsp d}{3}+4\multsp
{d^2}\Big)\multsp {\ka^4}+\bigg(\frac{244\multsp
d}{45}-\frac{40\multsp {d^2}}{3}+8\multsp {d^3}\bigg)\multsp
{\ka^6}\nonumber\\
&&+\bigg(\frac{1210\multsp d}{63}+\frac{24\multsp
{d^2}}{5}-40\multsp {d^3}+16\multsp {d^4}\bigg)\multsp {\ka^8}+\ldots
\end{eqnarray}
or, expanded in $w$,
\begin{equation}
\EA=
1+2\multsp d\multsp w+(-2\multsp d+4\multsp {d^2})\multsp
{w^2}+(2\multsp d-8\multsp {d^2}+8\multsp {d^3})\multsp
{w^3}+(26\multsp d-16\multsp {d^2}-24\multsp {d^3}+16\multsp
{d^4})\multsp {w^4}+\ldots
\end{equation}

\renewcommand{\timestamp}{Time-stamp: "2004-08-07 20:04:48 daboul"}

\section{The Full Series}\label{sec_series}

\sts The series for the $d$-dimensional hyper-cubic lattice to order
13 need to take into account 20724 graphs of up to 13 edges, and are
hence calculated using computers.  We use the graph data files
that were originally prepared for \cite{wan_91} by Wan \et, and have
since been used in many studies.  Programs were written, that use
these data files to compute the series as outlined in the previous
sections.  Details of the algorithms, including important efficiency
considerations, are presented in
\inthesis{the appendix.}{\cite{DaboulPhD}.}

Tables \ref{tab_bim} to \ref{tab_tri} show the resulting series in
full. In Tab.~\ref{tab_bim}, for the bimodal distribution, the
coefficients are given as exact fractions. For the other distributions
they were in part calculated using the data type {\em long double} in
C++ which limits their accuracy.  For comparison we used two different
processor architectures where this data type is represented in either 96 or 128
bits, and also compared part of the data with calculations done in
double precision (64 bits). Small rounding errors are obvious in most
numbers, but further investigation shows, that for coefficients large
in absolute value, the numerical accuracies become important.
Originally we had included in this work the {\em exponential random
  distribution} which is also addressed in \cite{camp94b}. This
distribution decays slower than the others and the resulting
coefficients become very large in absolute value, to a degree that
intermediate numbers either can not be presented in long double
variables or the rounding errors become so dominant that the highest
order coefficients come out completely wrong.  We have started to
calculate the series using arbitrary-precision numerical libraries,
but that work was not ready in time to be included here. We
exclude the exponential distribution from the present work, and for
the remaining series present the coefficients in as many digits as we
expect to be correct from the comparisons mentioned above.

From experience we know that small changes in the coefficients do not
influence the results obtained from series analysis. Hence the
numerical inaccuracies present in the power series should not
influence our final results.  For the bimodal distribution we
supplement coefficients for orders $x^{14}$ and $x^{15}$, which were
calculated using the non-free-end (NFE) technique and associated graph
data. In this technique by Harris~\cite{Harris82} the thermodynamic
functions under study are renormalized in such a way that the
contribution from a graph with at least one free end (i.e. a vertex
with only one incident edge) vanishes. This renormalization is
possible for the bimodal distribution~\cite{KleinAAHM91} but was not
obtained for the others. We use equations from
\cite{KleinAAHM91} for the NFE-expansion in $w$, but do not describe
the process here since the series can also be obtained directly by
variable transformation from $w$ to $x$, which indeed we use as a
consistency check.

Several checks are performed to assure the correctness of our series
expansions: The first is a complete recalculation of the corresponding
series in~\cite{KleinAAHM91} for the bimodal distribution, which shows
that the algorithm and its implementation are basically correct.

We mentioned earlier that after cumulant subtraction, a graph of $b$
edges has only terms of order $b$ and higher in the expansion
variable.  As an additional check we do the actual calculation of the
vanishing terms, track the maximal deviation from zero, and confirm
that this number is in the same range as the numerical rounding errors
observed elsewhere.

For a few sequences of coefficients we find, by examination of the
numerical values, what their exact value must be in general.  If we
denote by $a_{ij}$ the coefficient multiplying $x^i d^j$ we observe:
\begin{itemize}
\item For the bimodal distribution $a_{ii}=2^i$ and $a_{i,(i-1)}=-(5/6)\,
2^i\,(i-1)$.
\item For the Gaussian distribution $a_{ii}=2^i$ and $a_{i,(i-1)}=-(3/2)\,
2^i\,(i-1)$.
\item For the uniform distribution  $a_{ii}=(2/3)^i$ and $a_{i,(i-1)}=-(11/10)\,
(2/3)^i\,(i-1)$.
\item For the double-triangular distribution  $a_{ii}=1$ and
$a_{i,(i-1)}=-(17/18)\, (i-1)$.
\end{itemize}
Obviously this is no rigorous check from first principles, but if we
believe in the regularity and that we can at least calculate the first
few orders correctly, it adds confidence that no mistake was done at
higher orders and that the numerical errors are not exceedingly large.
A more comprehensive check for numerical rounding errors was already
mentioned above in this section.

\ifthenelse{\boolean{bo:aa}}{\linespread{1.3}}{}
\ifthenelse{1<0}{
\begin{table}
  \caption{Series for the Bimodal
    distribution. $\kex=(\Jcon/k_\mathrm{B}T)^2$ and $d$ is the spatial
    dimension of the hyper-cubic lattice.}\label{tab_bim}
  \begin{small}
    \begin{center}
                                                                            $+2$ & $x^{ 1}\, d^{ 1}$  & 
 & \\
\hline
\raisebox{-1.11ex}{\rule{0ex}{3.3ex}}                                                           $-\frac{10}{3}$ & $x^{ 2}\, d^{ 1}$  & 
                                                                      $+4$ & $x^{ 2}\, d^{ 2}$ \\
\hline
\raisebox{-1.11ex}{\rule{0ex}{3.3ex}}                                                         $+\frac{244}{45}$ & $x^{ 3}\, d^{ 1}$  & 
\raisebox{-1.11ex}{\rule{0ex}{3.3ex}}                                                           $-\frac{40}{3}$ & $x^{ 3}\, d^{ 2}$ \\
                                                                      $+8$ & $x^{ 3}\, d^{ 3}$  & 
 & \\
\hline
\raisebox{-1.11ex}{\rule{0ex}{3.3ex}}                                                        $+\frac{1210}{63}$ & $x^{ 4}\, d^{ 1}$  & 
\raisebox{-1.11ex}{\rule{0ex}{3.3ex}}                                                           $+\frac{24}{5}$ & $x^{ 4}\, d^{ 2}$ \\
                                                                     $-40$ & $x^{ 4}\, d^{ 3}$  & 
                                                                     $+16$ & $x^{ 4}\, d^{ 4}$ \\
\hline
\raisebox{-1.11ex}{\rule{0ex}{3.3ex}}                                                  $-\frac{2557316}{14175}$ & $x^{ 5}\, d^{ 1}$  & 
\raisebox{-1.11ex}{\rule{0ex}{3.3ex}}                                                      $+\frac{44480}{189}$ & $x^{ 5}\, d^{ 2}$ \\
\raisebox{-1.11ex}{\rule{0ex}{3.3ex}}                                                         $+\frac{296}{15}$ & $x^{ 5}\, d^{ 3}$  & 
\raisebox{-1.11ex}{\rule{0ex}{3.3ex}}                                                          $-\frac{320}{3}$ & $x^{ 5}\, d^{ 4}$ \\
                                                                     $+32$ & $x^{ 5}\, d^{ 5}$  & 
 & \\
\hline
\raisebox{-1.11ex}{\rule{0ex}{3.3ex}}                                                 $+\frac{15891824}{93555}$ & $x^{ 6}\, d^{ 1}$  & 
\raisebox{-1.11ex}{\rule{0ex}{3.3ex}}                                                  $-\frac{9373372}{14175}$ & $x^{ 6}\, d^{ 2}$ \\
\raisebox{-1.11ex}{\rule{0ex}{3.3ex}}                                                     $+\frac{111488}{189}$ & $x^{ 6}\, d^{ 3}$  & 
\raisebox{-1.11ex}{\rule{0ex}{3.3ex}}                                                        $+\frac{4688}{45}$ & $x^{ 6}\, d^{ 4}$ \\
\raisebox{-1.11ex}{\rule{0ex}{3.3ex}}                                                          $-\frac{800}{3}$ & $x^{ 6}\, d^{ 5}$  & 
                                                                     $+64$ & $x^{ 6}\, d^{ 6}$ \\
\hline
\raisebox{-1.11ex}{\rule{0ex}{3.3ex}}                                          $+\frac{190090194848}{42567525}$ & $x^{ 7}\, d^{ 1}$  & 
\raisebox{-1.11ex}{\rule{0ex}{3.3ex}}                                                  $-\frac{13641704}{2079}$ & $x^{ 7}\, d^{ 2}$ \\
\raisebox{-1.11ex}{\rule{0ex}{3.3ex}}                                                     $+\frac{645088}{945}$ & $x^{ 7}\, d^{ 3}$  & 
\raisebox{-1.11ex}{\rule{0ex}{3.3ex}}                                                     $+\frac{283264}{189}$ & $x^{ 7}\, d^{ 4}$ \\
\raisebox{-1.11ex}{\rule{0ex}{3.3ex}}                                                         $+\frac{1280}{3}$ & $x^{ 7}\, d^{ 5}$  & 
                                                                    $-640$ & $x^{ 7}\, d^{ 6}$ \\
                                                                    $+128$ & $x^{ 7}\, d^{ 7}$  & 
 & \\
\hline
\raisebox{-1.11ex}{\rule{0ex}{3.3ex}}                                         $-\frac{545049148646}{127702575}$ & $x^{ 8}\, d^{ 1}$  & 
\raisebox{-1.11ex}{\rule{0ex}{3.3ex}}                                        $+\frac{3811431542104}{212837625}$ & $x^{ 8}\, d^{ 2}$ \\
\raisebox{-1.11ex}{\rule{0ex}{3.3ex}}                                                  $-\frac{32973784}{2079}$ & $x^{ 8}\, d^{ 3}$  & 
\raisebox{-1.11ex}{\rule{0ex}{3.3ex}}                                                   $-\frac{7834448}{4725}$ & $x^{ 8}\, d^{ 4}$ \\
\raisebox{-1.11ex}{\rule{0ex}{3.3ex}}                                                      $+\frac{229856}{63}$ & $x^{ 8}\, d^{ 5}$  & 
\raisebox{-1.11ex}{\rule{0ex}{3.3ex}}                                                       $+\frac{22016}{15}$ & $x^{ 8}\, d^{ 6}$ \\
\raisebox{-1.11ex}{\rule{0ex}{3.3ex}}                                                         $-\frac{4480}{3}$ & $x^{ 8}\, d^{ 7}$  & 
                                                                    $+256$ & $x^{ 8}\, d^{ 8}$ \\
\hline
\raisebox{-1.11ex}{\rule{0ex}{3.3ex}}                                    $-\frac{2171514982687276}{8881133625}$ & $x^{ 9}\, d^{ 1}$  & 
\raisebox{-1.11ex}{\rule{0ex}{3.3ex}}                                        $+\frac{35779921623392}{76621545}$ & $x^{ 9}\, d^{ 2}$ \\
\raisebox{-1.11ex}{\rule{0ex}{3.3ex}}                                       $-\frac{32560925165624}{127702575}$ & $x^{ 9}\, d^{ 3}$  & 
\raisebox{-1.11ex}{\rule{0ex}{3.3ex}}                                               $+\frac{3151565216}{93555}$ & $x^{ 9}\, d^{ 4}$ \\
\raisebox{-1.11ex}{\rule{0ex}{3.3ex}}                                                  $-\frac{31525376}{2835}$ & $x^{ 9}\, d^{ 5}$  & 
\raisebox{-1.11ex}{\rule{0ex}{3.3ex}}                                                    $+\frac{1574144}{189}$ & $x^{ 9}\, d^{ 6}$ \\
\raisebox{-1.11ex}{\rule{0ex}{3.3ex}}                                                      $+\frac{203392}{45}$ & $x^{ 9}\, d^{ 7}$  & 
\raisebox{-1.11ex}{\rule{0ex}{3.3ex}}                                                        $-\frac{10240}{3}$ & $x^{ 9}\, d^{ 8}$ \\
                                                                    $+512$ & $x^{ 9}\, d^{ 9}$  & 
 & \\
\hline
\raisebox{-1.11ex}{\rule{0ex}{3.3ex}}                               $-\frac{202257782879679928}{1856156927625}$ & $x^{10}\, d^{ 1}$  & 
\raisebox{-1.11ex}{\rule{0ex}{3.3ex}}                                    $-\frac{3150596158319108}{7753370625}$ & $x^{10}\, d^{ 2}$ \\
\raisebox{-1.11ex}{\rule{0ex}{3.3ex}}                                        $+\frac{17497504604224}{18243225}$ & $x^{10}\, d^{ 3}$  & 
\raisebox{-1.11ex}{\rule{0ex}{3.3ex}}                                        $-\frac{44042569593584}{91216125}$ & $x^{10}\, d^{ 4}$ \\
\raisebox{-1.11ex}{\rule{0ex}{3.3ex}}                                                 $+\frac{255487168}{4455}$ & $x^{10}\, d^{ 5}$  & 
\raisebox{-1.11ex}{\rule{0ex}{3.3ex}}                                                   $-\frac{28463552}{675}$ & $x^{10}\, d^{ 6}$ \\
\raisebox{-1.11ex}{\rule{0ex}{3.3ex}}                                                      $+\frac{474880}{27}$ & $x^{10}\, d^{ 7}$  & 
\raisebox{-1.11ex}{\rule{0ex}{3.3ex}}                                                      $+\frac{193792}{15}$ & $x^{10}\, d^{ 8}$ \\
                                                                   $-7680$ & $x^{10}\, d^{ 9}$  & 
                                                                   $+1024$ & $x^{10}\, d^{10}$ \\
\hline
\raisebox{-1.11ex}{\rule{0ex}{3.3ex}}                       $+\frac{44286591649508625456608}{2143861251406875}$ & $x^{11}\, d^{ 1}$  & 
\raisebox{-1.11ex}{\rule{0ex}{3.3ex}}                              $-\frac{24516788251206488696}{519723939735}$ & $x^{11}\, d^{ 2}$ \\
\raisebox{-1.11ex}{\rule{0ex}{3.3ex}}                                      $+\frac{2975087273749088}{80405325}$ & $x^{11}\, d^{ 3}$  & 
\raisebox{-1.11ex}{\rule{0ex}{3.3ex}}                                       $-\frac{491320094394464}{42567525}$ & $x^{11}\, d^{ 4}$ \\
\raisebox{-1.11ex}{\rule{0ex}{3.3ex}}                                        $+\frac{14408475958592}{14189175}$ & $x^{11}\, d^{ 5}$  & 
\raisebox{-1.11ex}{\rule{0ex}{3.3ex}}                                                 $+\frac{753069824}{6237}$ & $x^{11}\, d^{ 6}$ \\
\raisebox{-1.11ex}{\rule{0ex}{3.3ex}}                                                 $-\frac{270251392}{2025}$ & $x^{11}\, d^{ 7}$  & 
\raisebox{-1.11ex}{\rule{0ex}{3.3ex}}                                                    $+\frac{6347776}{189}$ & $x^{11}\, d^{ 8}$ \\
\raisebox{-1.11ex}{\rule{0ex}{3.3ex}}                                                       $+\frac{175104}{5}$ & $x^{11}\, d^{ 9}$  & 
\raisebox{-1.11ex}{\rule{0ex}{3.3ex}}                                                        $-\frac{51200}{3}$ & $x^{11}\, d^{10}$ \\
                                                                   $+2048$ & $x^{11}\, d^{11}$  & 
 & \\
\hline
\raisebox{-1.11ex}{\rule{0ex}{3.3ex}}                      $+\frac{102687986431081211931032}{2275791174570375}$ & $x^{12}\, d^{ 1}$  & 
\raisebox{-1.11ex}{\rule{0ex}{3.3ex}}                       $-\frac{98234614240598344870804}{2143861251406875}$ & $x^{12}\, d^{ 2}$ \\
\raisebox{-1.11ex}{\rule{0ex}{3.3ex}}                                $-\frac{1631193202018689472}{39978764595}$ & $x^{12}\, d^{ 3}$  & 
\raisebox{-1.11ex}{\rule{0ex}{3.3ex}}                                  $+\frac{132365195242707824}{2170943775}$ & $x^{12}\, d^{ 4}$ \\
\raisebox{-1.11ex}{\rule{0ex}{3.3ex}}                                         $-\frac{36568729916000}{1702701}$ & $x^{12}\, d^{ 5}$  & 
\raisebox{-1.11ex}{\rule{0ex}{3.3ex}}                                          $+\frac{5544027735104}{2837835}$ & $x^{12}\, d^{ 6}$ \\
\raisebox{-1.11ex}{\rule{0ex}{3.3ex}}                                                   $+\frac{96053248}{297}$ & $x^{12}\, d^{ 7}$  & 
\raisebox{-1.11ex}{\rule{0ex}{3.3ex}}                                                    $-\frac{24126464}{63}$ & $x^{12}\, d^{ 8}$ \\
\raisebox{-1.11ex}{\rule{0ex}{3.3ex}}                                                     $+\frac{3447296}{63}$ & $x^{12}\, d^{ 9}$  & 
\raisebox{-1.11ex}{\rule{0ex}{3.3ex}}                                                       $+\frac{821248}{9}$ & $x^{12}\, d^{10}$ \\
\raisebox{-1.11ex}{\rule{0ex}{3.3ex}}                                                       $-\frac{112640}{3}$ & $x^{12}\, d^{11}$  & 
                                                                   $+4096$ & $x^{12}\, d^{12}$ \\
\hline
\raisebox{-1.11ex}{\rule{0ex}{3.3ex}}            $-\frac{110468581411293350924457444112}{48076088562799171875}$ & $x^{13}\, d^{ 1}$  & 
\raisebox{-1.11ex}{\rule{0ex}{3.3ex}}                        $+\frac{172781445528971814087368}{28988129795625}$ & $x^{13}\, d^{ 2}$ \\
\raisebox{-1.11ex}{\rule{0ex}{3.3ex}}                      $-\frac{2939279178242203350187328}{510443155096875}$ & $x^{13}\, d^{ 3}$  & 
\raisebox{-1.11ex}{\rule{0ex}{3.3ex}}                           $+\frac{4778773293081621239776}{1856156927625}$ & $x^{13}\, d^{ 4}$ \\
\raisebox{-1.11ex}{\rule{0ex}{3.3ex}}                                     $-\frac{23565892697470112}{46990125}$ & $x^{13}\, d^{ 5}$  & 
\raisebox{-1.11ex}{\rule{0ex}{3.3ex}}                                           $+\frac{5353769766272}{289575}$ & $x^{13}\, d^{ 6}$ \\
\raisebox{-1.11ex}{\rule{0ex}{3.3ex}}                                        $+\frac{39918090277888}{10135125}$ & $x^{13}\, d^{ 7}$  & 
\raisebox{-1.11ex}{\rule{0ex}{3.3ex}}                                                  $+\frac{688011776}{693}$ & $x^{13}\, d^{ 8}$ \\
\raisebox{-1.11ex}{\rule{0ex}{3.3ex}}                                                  $-\frac{179936768}{175}$ & $x^{13}\, d^{ 9}$  & 
\raisebox{-1.11ex}{\rule{0ex}{3.3ex}}                                                     $+\frac{1286144}{21}$ & $x^{13}\, d^{10}$ \\
\raisebox{-1.11ex}{\rule{0ex}{3.3ex}}                                                      $+\frac{1153024}{5}$ & $x^{13}\, d^{11}$  & 
                                                                  $-81920$ & $x^{13}\, d^{12}$ \\
                                                                   $+8192$ & $x^{13}\, d^{13}$  & 
 & \\
\hline
\raisebox{-1.11ex}{\rule{0ex}{3.3ex}}            $-\frac{305463146085574972582952872664}{33041384503160158125}$ & $x^{14}\, d^{ 1}$  & 
\raisebox{-1.11ex}{\rule{0ex}{3.3ex}}        $+\frac{50334647726118167100558221079016}{3028793579456347828125}$ & $x^{14}\, d^{ 2}$ \\
\raisebox{-1.11ex}{\rule{0ex}{3.3ex}}                 $-\frac{824015715606745029679283608}{147926426347074375}$ & $x^{14}\, d^{ 3}$  & 
\raisebox{-1.11ex}{\rule{0ex}{3.3ex}}                  $-\frac{157096153765758568475862928}{32157918771103125}$ & $x^{14}\, d^{ 4}$ \\
\raisebox{-1.11ex}{\rule{0ex}{3.3ex}}                          $+\frac{30648160725228437701792}{7795859096025}$ & $x^{14}\, d^{ 5}$  & 
\raisebox{-1.11ex}{\rule{0ex}{3.3ex}}                                 $-\frac{2624211856887202496}{2960377875}$ & $x^{14}\, d^{ 6}$ \\
\raisebox{-1.11ex}{\rule{0ex}{3.3ex}}                                         $+\frac{43529789374208}{1563705}$ & $x^{14}\, d^{ 7}$  & 
\raisebox{-1.11ex}{\rule{0ex}{3.3ex}}                                     $+\frac{1037253983072768}{127702575}$ & $x^{14}\, d^{ 8}$ \\
\raisebox{-1.11ex}{\rule{0ex}{3.3ex}}                                                 $+\frac{1206767104}{385}$ & $x^{14}\, d^{ 9}$  & 
\raisebox{-1.11ex}{\rule{0ex}{3.3ex}}                                                 $-\frac{1487067136}{567}$ & $x^{14}\, d^{10}$ \\
\raisebox{-1.11ex}{\rule{0ex}{3.3ex}}                                                    $-\frac{5660672}{189}$ & $x^{14}\, d^{11}$  & 
\raisebox{-1.11ex}{\rule{0ex}{3.3ex}}                                                     $+\frac{8531968}{15}$ & $x^{14}\, d^{12}$ \\
\raisebox{-1.11ex}{\rule{0ex}{3.3ex}}                                                       $-\frac{532480}{3}$ & $x^{14}\, d^{13}$  & 
                                                                  $+16384$ & $x^{14}\, d^{14}$ \\
\hline
\raisebox{-1.11ex}{\rule{0ex}{3.3ex}}$+\frac{1293826468733333294113597991611697536}{3952575621190533915703125}$ & $x^{15}\, d^{ 1}$  & 
\raisebox{-1.11ex}{\rule{0ex}{3.3ex}}        $-\frac{338533911194409672724356315499808}{363455229534761739375}$ & $x^{15}\, d^{ 2}$ \\
\raisebox{-1.11ex}{\rule{0ex}{3.3ex}}      $+\frac{3142897456730380314538798434325456}{3028793579456347828125}$ & $x^{15}\, d^{ 3}$  & 
\raisebox{-1.11ex}{\rule{0ex}{3.3ex}}              $-\frac{257903955006958171637687615776}{443779279041223125}$ & $x^{15}\, d^{ 4}$ \\
\raisebox{-1.11ex}{\rule{0ex}{3.3ex}}                 $+\frac{5449914292983317242598754368}{32157918771103125}$ & $x^{15}\, d^{ 5}$  & 
\raisebox{-1.11ex}{\rule{0ex}{3.3ex}}                          $-\frac{67310752760322121480064}{2998407344625}$ & $x^{15}\, d^{ 6}$ \\
\raisebox{-1.11ex}{\rule{0ex}{3.3ex}}                               $+\frac{11493150389809720832}{23260111875}$ & $x^{15}\, d^{ 7}$  & 
\raisebox{-1.11ex}{\rule{0ex}{3.3ex}}                                     $+\frac{4618407513611776}{127702575}$ & $x^{15}\, d^{ 8}$ \\
\raisebox{-1.11ex}{\rule{0ex}{3.3ex}}                                     $+\frac{3526208986206208}{212837625}$ & $x^{15}\, d^{ 9}$  & 
\raisebox{-1.11ex}{\rule{0ex}{3.3ex}}                                             $+\frac{904493910016}{93555}$ & $x^{15}\, d^{10}$ \\
\raisebox{-1.11ex}{\rule{0ex}{3.3ex}}                                              $-\frac{90834735104}{14175}$ & $x^{15}\, d^{11}$  & 
\raisebox{-1.11ex}{\rule{0ex}{3.3ex}}                                                   $-\frac{93716480}{189}$ & $x^{15}\, d^{12}$ \\
\raisebox{-1.11ex}{\rule{0ex}{3.3ex}}                                                    $+\frac{61898752}{45}$ & $x^{15}\, d^{13}$  & 
\raisebox{-1.11ex}{\rule{0ex}{3.3ex}}                                                      $-\frac{1146880}{3}$ & $x^{15}\, d^{14}$ \\
                                                                  $+32768$ & $x^{15}\, d^{15}$  & 
 & \\
\hline
    \end{center}
  \end{small}
\end{table}
}{
\setlength{\LTcapwidth}{\textwidth}
\begin{longtable}{@{}r@{$\;$}l@{}r@{$\;$}l@{}}
\caption{Series for the Bimodal distribution on the $d$-dimensional
 hyper-cubic lattice and for $\kex=(\Jcon/k_\mathrm{B}T)^2$.\label{tab_bim}}\\
  \hline
  \multicolumn{4}{c}{Terms of the series. $\EA=1+\ldots$ }\\
  \hline
  \hline
  \endfirsthead
  \caption{Series for the Bimodal distribution. (continued)}\\
  \hline
  \multicolumn{4}{c}{Terms of the series.}\\ 
  \hline
  \hline
  \endhead
  \hline
  \endfoot
  \hline
  \endlastfoot
\end{longtable}
}
\begin{table}
  \caption{Series for the Gaussian distribution on the $d$-dimensional
 hyper-cubic lattice and for $\kex=(\Jcon/k_\mathrm{B}T)^2$.}\label{tab_gau}
  \begin{center}
    \begin{small}
      \inthesis{\makebox[\textwidth][c]{%
          \begin{minipage}{1.35\textwidth}
            \begin{center}
  \begin{tabular}{@{}r@{$\;$}l@{}r@{$\;$}l@{}r@{$\;$}l@{}}
    \hline
    \multicolumn{6}{c}{Terms of the series. $\EA=1+\ldots$ } \\ 
    \hline
    \hline
 $+2$                    & $x^{ 1}\,d^{ 1}$  &  & &  &                                                                                  \\
  \hline
 $-6$                    & $x^{ 2}\,d^{ 1}$  &  $+4$ & $x^{ 2}\,d^{ 2}$  &  &                                                           \\
  \hline
 $+21.33333333333333333$ & $x^{ 3}\,d^{ 1}$  &  $-24$ & $x^{ 3}\,d^{ 2}$  &  $+8$ & $x^{ 3}\,d^{ 3}$                                    \\
  \hline
 $-57.9999999999999999$  & $x^{ 4}\,d^{ 1}$  &  $+93.333333333333333$ & $x^{ 4}\,d^{ 2}$  &  $-71.999999999999999$ & $x^{ 4}\,d^{ 3}$   \\
 $+15.999999999999999$   & $x^{ 4}\,d^{ 4}$  &  & &  &                                                                                  \\
  \hline
 $+20.266666666666666$   & $x^{ 5}\,d^{ 1}$  &  $-119.999999999999999$ & $x^{ 5}\,d^{ 2}$  &  $+359.999999999999999$ & $x^{ 5}\,d^{ 3}$ \\
 $-192$                  & $x^{ 5}\,d^{ 4}$  &  $+32$ & $x^{ 5}\,d^{ 5}$  &  &                                                          \\
  \hline
 $+558.4000000000000$    & $x^{ 6}\,d^{ 1}$  &  $-934.488888888888$ & $x^{ 6}\,d^{ 2}$  &  $-991.999999999999$ & $x^{ 6}\,d^{ 3}$       \\
 $+1210.666666666666$    & $x^{ 6}\,d^{ 4}$  &  $-479.999999999999$ & $x^{ 6}\,d^{ 5}$  &  $+63.999999999999$ & $x^{ 6}\,d^{ 6}$        \\
  \hline
 $+1000.02539682539$     & $x^{ 7}\,d^{ 1}$  &  $+2374.39999999999$ & $x^{ 7}\,d^{ 2}$  &  $+2408.5333333333$ & $x^{ 7}\,d^{ 3}$        \\
 $-4704.0000000000$      & $x^{ 7}\,d^{ 4}$  &  $+3690.6666666666$ & $x^{ 7}\,d^{ 5}$  &  $-1152.0000000000$ & $x^{ 7}\,d^{ 6}$         \\
 $+128.00000000000$      & $x^{ 7}\,d^{ 7}$  &  & &  &                                                                                  \\
  \hline
 $-31435.219047619$      & $x^{ 8}\,d^{ 1}$  &  $+21442.41269841$ & $x^{ 8}\,d^{ 2}$  &  $-20110.400000000$ & $x^{ 8}\,d^{ 3}$          \\
 $+12771.200000000$      & $x^{ 8}\,d^{ 4}$  &  $-18143.999999999$ & $x^{ 8}\,d^{ 5}$  &  $+10495.999999999$ & $x^{ 8}\,d^{ 6}$         \\
 $-2687.999999999$       & $x^{ 8}\,d^{ 7}$  &  $+255.9999999999$ & $x^{ 8}\,d^{ 8}$  &  &                                              \\
  \hline
 $-20387.53298060$       & $x^{ 9}\,d^{ 1}$  &  $+294403.1746032$ & $x^{ 9}\,d^{ 2}$  &  $-95832.1693122$ & $x^{ 9}\,d^{ 3}$            \\
 $+21440.0000000$        & $x^{ 9}\,d^{ 4}$  &  $+61560.888888$ & $x^{ 9}\,d^{ 5}$  &  $-62080.000000$ & $x^{ 9}\,d^{ 6}$               \\
 $+28373.333333$         & $x^{ 9}\,d^{ 7}$  &  $-6144.0000000$ & $x^{ 9}\,d^{ 8}$  &  $+512.00000000$ & $x^{ 9}\,d^{ 9}$               \\
  \hline
 $+2051214.7843386$      & $x^{10}\,d^{ 1}$  &  $-6192978.8227$ & $x^{10}\,d^{ 2}$  &  $+3181210.20952$ & $x^{10}\,d^{ 3}$              \\
 $-1079716.757669$       & $x^{10}\,d^{ 4}$  &  $-48661.333333$ & $x^{10}\,d^{ 5}$  &  $+257267.199999$ & $x^{10}\,d^{ 6}$              \\
 $-195840.00000$         & $x^{10}\,d^{ 7}$  &  $+73813.33333$ & $x^{10}\,d^{ 8}$  &  $-13824.00000$ & $x^{10}\,d^{ 9}$                 \\
 $+1024.000000$          & $x^{10}\,d^{10}$  &  & &  &                                                                                  \\
  \hline
 $-2725463.2041$         & $x^{11}\,d^{ 1}$  &  $+18390206.21$ & $x^{11}\,d^{ 2}$  &  $+2845075.7280$ & $x^{11}\,d^{ 3}$                \\
 $+798591.1877$          & $x^{11}\,d^{ 4}$  &  $-340330.5315$ & $x^{11}\,d^{ 5}$  &  $-554547.199$ & $x^{11}\,d^{ 6}$                  \\
 $+961186.1329$          & $x^{11}\,d^{ 7}$  &  $-582143.999$ & $x^{11}\,d^{ 8}$  &  $+186367.999$ & $x^{11}\,d^{ 9}$                   \\
 $-30719.999$            & $x^{11}\,d^{10}$  &  $+2047.9999$ & $x^{11}\,d^{11}$  &  &                                                   \\
  \hline
 $-52046138.6$           & $x^{12}\,d^{ 1}$  &  $+64425096.1$ & $x^{12}\,d^{ 2}$  &  $-314627076.1$ & $x^{12}\,d^{ 3}$                  \\
 $+126856674.2$          & $x^{12}\,d^{ 4}$  &  $-20441160.8$ & $x^{12}\,d^{ 5}$  &  $-520014.1$ & $x^{12}\,d^{ 6}$                     \\
 $-3007795.2$            & $x^{12}\,d^{ 7}$  &  $+3301421$ & $x^{12}\,d^{ 8}$  &  $-1653248$ & $x^{12}\,d^{ 9}$                         \\
 $+459434$               & $x^{12}\,d^{10}$  &  $-67584.0$ & $x^{12}\,d^{11}$  &  $+4096.00$ & $x^{12}\,d^{12}$                         \\
  \hline
 $-2052218007$           & $x^{13}\,d^{ 1}$  &  $+563481178 e 1$ & $x^{13}\,d^{ 2}$  &  $-20936942 e 2 $ & $x^{13}\,d^{ 3}$             \\
 $+846300080$            & $x^{13}\,d^{ 4}$  &  $-15546990 e 1$ & $x^{13}\,d^{ 5}$  &  $+318945 e 2$ & $x^{13}\,d^{ 6}$                 \\
 $+2724596$              & $x^{13}\,d^{ 7}$  &  $-1300363 e 1$ & $x^{13}\,d^{ 8}$  &  $+1063195 e 1$ & $x^{13}\,d^{ 9}$                 \\
 $-45281 e 2$            & $x^{13}\,d^{10}$  &  $+1110 e 3$ & $x^{13}\,d^{11}$  &  $-147456$ & $x^{13}\,d^{12}$                         \\
 $+8192.0$               & $x^{13}\,d^{13}$  &  & &  &                                                                                  \\
  \hline
  \end{tabular}

            \end{center}
          \end{minipage}}
      }{}
    \end{small}
  \end{center}
\end{table}
\begin{table}
  \caption{Series for the Uniform distribution on the $d$-dimensional
 hyper-cubic lattice and for $\kex=(\Jcon/k_\mathrm{B}T)^2$.}\label{tab_uni}
  \begin{center}
    \begin{small}
      \inthesis{\makebox[\textwidth][c]{%
          \begin{minipage}{1.35\textwidth}
            \begin{center}
  \begin{tabular}{@{}r@{$\;$}l@{}r@{$\;$}l@{}r@{$\;$}l@{}}
    \hline
    \multicolumn{6}{c}{Terms of the series. $\EA=1+\ldots$ } \\ 
    \hline
    \hline
                     $+0.6666666666666666666$ & $x^{ 1}\,d^{ 1}$  &  & &  & \\
    \hline
                     $-0.488888888888888888$ & $x^{ 2}\,d^{ 1}$  &                     $+0.4444444444444444444$ & $x^{ 2}\,d^{ 2}$  &  & \\
    \hline
                     $+0.359788359788359788$ & $x^{ 3}\,d^{ 1}$  &                     $-0.65185185185185185$ & $x^{ 3}\,d^{ 2}$  &                     $+0.296296296296296296$ & $x^{ 3}\,d^{ 3}$  \\
    \hline
                     $+0.08084656084656084$ & $x^{ 4}\,d^{ 1}$  &                     $+0.373051146384479718$ & $x^{ 4}\,d^{ 2}$  &                     $-0.65185185185185185$ & $x^{ 4}\,d^{ 3}$  \\
                     $+0.19753086419753086$ & $x^{ 4}\,d^{ 4}$  &  & &  & \\
    \hline
                     $-0.89147987814654481$ & $x^{ 5}\,d^{ 1}$  &                     $+0.84242210464432686$ & $x^{ 5}\,d^{ 2}$  &                     $+0.49683715461493239$ & $x^{ 5}\,d^{ 3}$  \\
                     $-0.57942386831275720$ & $x^{ 5}\,d^{ 4}$  &                     $+0.13168724279835391$ & $x^{ 5}\,d^{ 5}$  &  & \\
    \hline
                     $+1.0637968983789089$ & $x^{ 6}\,d^{ 1}$  &                     $-1.862678750213141$ & $x^{ 6}\,d^{ 2}$  &                     $+0.59106251224769744$ & $x^{ 6}\,d^{ 3}$  \\
                     $+0.60287673917303544$ & $x^{ 6}\,d^{ 4}$  &                     $-0.48285322359396431$ & $x^{ 6}\,d^{ 5}$  &                     $+0.08779149519890260$ & $x^{ 6}\,d^{ 6}$  \\
    \hline
                     $+1.255336691527167$ & $x^{ 7}\,d^{ 1}$  &                     $-1.283045789797112$ & $x^{ 7}\,d^{ 2}$  &                     $-0.625225076837070$ & $x^{ 7}\,d^{ 3}$  \\
                     $+0.3268519171729049$ & $x^{ 7}\,d^{ 4}$  &                     $+0.6538376118623032$ & $x^{ 7}\,d^{ 5}$  &                     $-0.3862825788751715$ & $x^{ 7}\,d^{ 6}$  \\
                     $+0.0585276634659350$ & $x^{ 7}\,d^{ 7}$  &  & &  & \\
    \hline
                     $-2.764322363362706$ & $x^{ 8}\,d^{ 1}$  &                     $+5.381812025290449$ & $x^{ 8}\,d^{ 2}$  &                     $-1.970220360622241$ & $x^{ 8}\,d^{ 3}$  \\
                     $-1.100189850013248$ & $x^{ 8}\,d^{ 4}$  &                     $+0.06329348749101851$ & $x^{ 8}\,d^{ 5}$  &                     $+0.6510505802686874$ & $x^{ 8}\,d^{ 6}$  \\
                     $-0.300442005791800$ & $x^{ 8}\,d^{ 7}$  &                     $+0.03901844231062340$ & $x^{ 8}\,d^{ 8}$  &  & \\
    \hline
                     $-9.85782657798808$ & $x^{ 9}\,d^{ 1}$  &                     $+17.54278833330664$ & $x^{ 9}\,d^{ 2}$  &                     $-9.714423500532811$ & $x^{ 9}\,d^{ 3}$  \\
                     $+3.026748425908237$ & $x^{ 9}\,d^{ 4}$  &                     $-1.234630286704749$ & $x^{ 9}\,d^{ 5}$  &                     $-0.168708312466884$ & $x^{ 9}\,d^{ 6}$  \\
                     $+0.608947822994459$ & $x^{ 9}\,d^{ 7}$  &                     $-0.22890819488898$ & $x^{ 9}\,d^{ 8}$  &                     $+0.0260122948737485$ & $x^{ 9}\,d^{ 9}$  \\
    \hline
                     $+15.5044273681052$ & $x^{10}\,d^{ 1}$  &                     $-38.7783044835667$ & $x^{10}\,d^{ 2}$  &                     $+33.5641685051471$ & $x^{10}\,d^{ 3}$  \\
                     $-12.0316553472705$ & $x^{10}\,d^{ 4}$  &                     $+2.84533673882951$ & $x^{10}\,d^{ 5}$  &                     $-1.14514652844366$ & $x^{10}\,d^{ 6}$  \\
                     $-0.348044505410769$ & $x^{10}\,d^{ 7}$  &                     $+0.543557868404698$ & $x^{10}\,d^{ 8}$  &                     $-0.171681146166746$ & $x^{10}\,d^{ 9}$  \\
                     $+0.0173415299158332$ & $x^{10}\,d^{10}$  &  & &  & \\
    \hline
                     $+108.085493498972$ & $x^{11}\,d^{ 1}$  &                     $-232.015426078230$ & $x^{11}\,d^{ 2}$  &                     $+168.045218530949$ & $x^{11}\,d^{ 3}$  \\
                     $-48.1504957504608$ & $x^{11}\,d^{ 4}$  &                     $+2.24480229556376$ & $x^{11}\,d^{ 5}$  &                     $+2.83058482344332$ & $x^{11}\,d^{ 6}$  \\
                     $-0.924877616717460$ & $x^{11}\,d^{ 7}$  &                     $-0.467778685821369$ & $x^{11}\,d^{ 8}$  &                     $+0.468089181785302$ & $x^{11}\,d^{ 9}$  \\
                     $-0.127171219382772$ & $x^{11}\,d^{10}$  &                     $+0.0115610199438887$ & $x^{11}\,d^{11}$  &  & \\
    \hline
                     $-142.76066452364$ & $x^{12}\,d^{ 1}$  &                     $+418.50354734964$ & $x^{12}\,d^{ 2}$  &                     $-458.98712418598$ & $x^{12}\,d^{ 3}$  \\
                     $+231.20946538453$ & $x^{12}\,d^{ 4}$  &                     $-49.940080322749$ & $x^{12}\,d^{ 5}$  &                     $+0.060628792921663$ & $x^{12}\,d^{ 6}$  \\
                     $+2.7869356691639$ & $x^{12}\,d^{ 7}$  &                     $-0.64727598233428$ & $x^{12}\,d^{ 8}$  &                     $-0.53174415759476$ & $x^{12}\,d^{ 9}$  \\
                     $+0.39186352362185$ & $x^{12}\,d^{10}$  &                     $-0.093258894214093$ & $x^{12}\,d^{11}$  &                     $+0.0077073466292662$ & $x^{12}\,d^{12}$  \\
    \hline
                     $-1391.033864136$ & $x^{13}\,d^{ 1}$  &                     $+3401.571257928$ & $x^{13}\,d^{ 2}$  &                     $-3040.190268119$ & $x^{13}\,d^{ 3}$  \\
                     $+1234.36058382$ & $x^{13}\,d^{ 4}$  &                     $-211.0538211516$ & $x^{13}\,d^{ 5}$  &                     $+5.814822817330$ & $x^{13}\,d^{ 6}$  \\
                     $-1.45093243110$ & $x^{13}\,d^{ 7}$  &                     $+2.63953125141$ & $x^{13}\,d^{ 8}$  &                     $-0.365539203648$ & $x^{13}\,d^{ 9}$  \\
                     $-0.5497466842$ & $x^{13}\,d^{10}$  &                     $+0.32066232142$ & $x^{13}\,d^{11}$  &                     $-0.0678246503365$ & $x^{13}\,d^{12}$  \\
                     $+0.005138231086$ & $x^{13}\,d^{13}$  &  & &  & \\
    \hline
  \end{tabular}

            \end{center}
          \end{minipage}}
      }{}
    \end{small}
  \end{center}
\end{table}
\begin{table}
  \caption{Series for the Double-Triangular distribution on the $d$-dimensional
 hyper-cubic lattice and for $\kex=(\Jcon/k_\mathrm{B}T)^2$.}\label{tab_tri}
  \begin{center}
    \begin{small}
      \inthesis{\makebox[\textwidth][c]{%
          \begin{minipage}{1.35\textwidth}
            \begin{center}
  \begin{tabular}{@{}r@{$\;$}l@{}r@{$\;$}l@{}r@{$\;$}l@{}}
    \hline
    \multicolumn{6}{c}{Terms of the series. $\EA=1+\ldots$ } \\ 
    \hline
    \hline
                     $+1$ & $x^{ 1}\,d^{ 1}$  &  & &  & \\
    \hline
                     $-0.944444444444444444$ & $x^{ 2}\,d^{ 1}$  &                     $+1$ & $x^{ 2}\,d^{ 2}$  &  & \\
    \hline
                     $+0.883333333333333333$ & $x^{ 3}\,d^{ 1}$  &                     $-1.88888888888888888$ & $x^{ 3}\,d^{ 2}$  &                     $+0.999999999999999999$ & $x^{ 3}\,d^{ 3}$  \\
    \hline
                     $+0.925282186948853615$ & $x^{ 4}\,d^{ 1}$  &                     $+0.908641975308641977$ & $x^{ 4}\,d^{ 2}$  &                     $-2.83333333333333333$ & $x^{ 4}\,d^{ 3}$  \\
                     $+1.00000000000000000$ & $x^{ 4}\,d^{ 4}$  &  & &  & \\
    \hline
                     $-6.21639476778365667$ & $x^{ 5}\,d^{ 1}$  &                     $+7.16815696649029982$ & $x^{ 5}\,d^{ 2}$  &                     $+1.82592592592592591$ & $x^{ 5}\,d^{ 3}$  \\
                     $-3.77777777777777777$ & $x^{ 5}\,d^{ 4}$  &                     $+0.99999999999999999$ & $x^{ 5}\,d^{ 5}$  &  & \\
    \hline
                     $+6.3863053214177552$ & $x^{ 6}\,d^{ 1}$  &                     $-15.004896629433666$ & $x^{ 6}\,d^{ 2}$  &                     $+8.705647658240250$ & $x^{ 6}\,d^{ 3}$  \\
                     $+3.63518518518518$ & $x^{ 6}\,d^{ 4}$  &                     $-4.722222222222222$ & $x^{ 6}\,d^{ 5}$  &                     $+0.9999999999999999$ & $x^{ 6}\,d^{ 6}$  \\
    \hline
                     $+31.28590186111328$ & $x^{ 7}\,d^{ 1}$  &                     $-41.38286394727488$ & $x^{ 7}\,d^{ 2}$  &                     $-1.768129041740152$ & $x^{ 7}\,d^{ 3}$  \\
                     $+10.19533313737017$ & $x^{ 7}\,d^{ 4}$  &                     $+6.336419753086420$ & $x^{ 7}\,d^{ 5}$  &                     $-5.666666666666666$ & $x^{ 7}\,d^{ 6}$  \\
                     $+1.000000000000000$ & $x^{ 7}\,d^{ 7}$  &  & &  & \\
    \hline
                     $-42.9465436061225$ & $x^{ 8}\,d^{ 1}$  &                     $+105.962396480164$ & $x^{ 8}\,d^{ 2}$  &                     $-61.3217767001924$ & $x^{ 8}\,d^{ 3}$  \\
                     $-16.807385633723$ & $x^{ 8}\,d^{ 4}$  &                     $+10.7947922790515$ & $x^{ 8}\,d^{ 5}$  &                     $+9.92962962962962$ & $x^{ 8}\,d^{ 6}$  \\
                     $-6.61111111111111$ & $x^{ 8}\,d^{ 7}$  &                     $+0.999999999999999$ & $x^{ 8}\,d^{ 8}$  &  & \\
    \hline
                     $-447.38013758662$ & $x^{ 9}\,d^{ 1}$  &                     $+826.325405622926$ & $x^{ 9}\,d^{ 2}$  &                     $-441.151661966652$ & $x^{ 9}\,d^{ 3}$  \\
                     $+80.2235926358573$ & $x^{ 9}\,d^{ 4}$  &                     $-35.5380623053978$ & $x^{ 9}\,d^{ 5}$  &                     $+9.66160395845584$ & $x^{ 9}\,d^{ 6}$  \\
                     $+14.4148148148148$ & $x^{ 9}\,d^{ 7}$  &                     $-7.55555555555555$ & $x^{ 9}\,d^{ 8}$  &                     $+1.00000000000000$ & $x^{ 9}\,d^{ 9}$  \\
    \hline
                     $+331.676191097529$ & $x^{10}\,d^{ 1}$  &                     $-1222.22773822861$ & $x^{10}\,d^{ 2}$  &                     $+1383.83407091038$ & $x^{10}\,d^{ 3}$  \\
                     $-545.353553191950$ & $x^{10}\,d^{ 4}$  &                     $+91.7181421721279$ & $x^{10}\,d^{ 5}$  &                     $-57.8924350055523$ & $x^{10}\,d^{ 6}$  \\
                     $+5.95334705075442$ & $x^{10}\,d^{ 7}$  &                     $+19.7919753086420$ & $x^{10}\,d^{ 8}$  &                     $-8.5000000000000$ & $x^{10}\,d^{ 9}$  \\
                     $+1.00000000000000$ & $x^{10}\,d^{10}$  &  & &  & \\
    \hline
                     $+10066.680607773$ & $x^{11}\,d^{ 1}$  &                     $-22331.632875336$ & $x^{11}\,d^{ 2}$  &                     $+16898.722229737$ & $x^{11}\,d^{ 3}$  \\
                     $-5104.9579819064$ & $x^{11}\,d^{ 4}$  &                     $+412.40450343047$ & $x^{11}\,d^{ 5}$  &                     $+125.3464089002$ & $x^{11}\,d^{ 6}$  \\
                     $-83.007159731749$ & $x^{11}\,d^{ 7}$  &                     $-1.1723995688801$ & $x^{11}\,d^{ 8}$  &                     $+26.061111111110$ & $x^{11}\,d^{ 9}$  \\
                     $-9.444444444444$ & $x^{11}\,d^{10}$  &                     $+0.99999999999998$ & $x^{11}\,d^{11}$  &  & \\
    \hline
                     $-1250.8803969425$ & $x^{12}\,d^{ 1}$  &                     $+16566.72572040$ & $x^{12}\,d^{ 2}$  &                     $-31549.37183747$ & $x^{12}\,d^{ 3}$  \\
                     $+21664.50242641$ & $x^{12}\,d^{ 4}$  &                     $-5879.76957662$ & $x^{12}\,d^{ 5}$  &                     $+360.0325939430$ & $x^{12}\,d^{ 6}$  \\
                     $+186.7090665165$ & $x^{12}\,d^{ 7}$  &                     $-109.223272530$ & $x^{12}\,d^{ 8}$  &                     $-12.5580570252$ & $x^{12}\,d^{ 9}$  \\
                     $+33.22222222221$ & $x^{12}\,d^{10}$  &                     $-10.38888888888$ & $x^{12}\,d^{11}$  &                     $+0.999999999999$ & $x^{12}\,d^{12}$  \\
    \hline
                     $-288282.12879$ & $x^{13}\,d^{ 1}$  &                     $+726711.250361$ & $x^{13}\,d^{ 2}$  &                     $-676798.593987$ & $x^{13}\,d^{ 3}$  \\
                     $+289743.600575$ & $x^{13}\,d^{ 4}$  &                     $-53400.7324666$ & $x^{13}\,d^{ 5}$  &                     $+1584.74112371$ & $x^{13}\,d^{ 6}$  \\
                     $+292.731945980$ & $x^{13}\,d^{ 7}$  &                     $+281.321506576$ & $x^{13}\,d^{ 8}$  &                     $-134.086189495$ & $x^{13}\,d^{ 9}$  \\
                     $-29.046046443$ & $x^{13}\,d^{10}$  &                     $+41.2753086420$ & $x^{13}\,d^{11}$  &                     $-11.3333333333$ & $x^{13}\,d^{12}$  \\
                     $+1.00000000000$ & $x^{13}\,d^{13}$  &  & &  & \\
    \hline
  \end{tabular}
 %
            \end{center}
          \end{minipage}}
      }{} %
    \end{small}
  \end{center}
\end{table}
\ifthenelse{\boolean{bo:aa}}{\linespread{1.6}}{}

\renewcommand{\timestamp}{Time-stamp: "2004-05-05 09:48:51 daboul"}
\section{Analysis of the Series}\label{sec:analysis}

\sts Our analysis uses the \dlog\ method~\cite{gamm73} and the methods
M1 and M2~\cite{AdlerMP82,adle93}.  Each of these is combined with
Euler-transformations for improved results.  For each series, our main
goal is to obtain the critical value $x_c$ and the critical exponent
$\gamma$ which describe the power law divergence, as in
\begin{equation}
  \typeout{repeated eq:scaling-form}
  \EA\approx A(\pq - x)^{-\gamma}(1+B(\pq - x)^{\Delta_1}).
\end{equation}

The series analysis is done for a fixed dimension at a time.  We
present our results for dimensions 7 and 8 above the upper critical
dimension and for 5 and 4 below it.  We also attempted an analysis in
the physical dimension 3 but the results are not conclusive.

At the upper critical dimension $d_{\mathrm{c}}=6$ the corrections to
scaling become logarithmic and there one expects the general form
\begin{equation}
  \label{eq:log-scaling}
  \EA(x)\approx A(x_\mathrm{c}-x)^{-\gamma} |\ln(x_\mathrm{c}-x)|^\theta\,.
\end{equation}
Instead of M1 and M2, one can apply a modified method to take such
corrections into account. This was pursued in \cite{KleinAAHM91}, for
the Bimodal distribution, but the authors reported poor convergence
already for that case.  Given that our series for the other
distributions are more problematic, we did not attempt a detailed
analysis in $d=6$.

It is generally observed in series analysis, that for a given order of
expansion, a series behaves better, the higher the dimension.  That is
also the case in the study at hand.  Qualitatively it is understood by
the fast increase of the embedding constants with increasing
dimension.  Thus a much larger number of lattice configurations
contributes to the higher dimensional series, allowing it to capture
more of the underlying Physics.

\subsection{\dlog\ Analysis}

The \dlog\ method is one of the most common methods for the asymptotic
analysis of power series. One calculates Pad\'e approximants to the
logarithmic derivative of the series and obtains estimates for the
critical value $x_c$ of the expansion variable $x$ (the threshold) and
for the critical exponent $\gamma$ from their real first order
poles and the corresponding residues.  We also refer to the
pole-residue pairs as data-points since we often plot them in diagrams
of residues versus poles.

Many series point to singularities other than those representing the
physical critical point.  They are observed in the \dlog\ analysis of the
original series and, depending on their strength and location in the
complex plane, hamper convergence of the data points.  This effect
appears to be strongest when an extra singularity is on the negative
real axis closer to the origin than the physical one.  Application of
an Euler-transformation into the new variable $z= x_n\ x/(x_n-x)$,
with $x_n$ at or close to the disturbing singularity, usually improves
the behavior of the transformed series.

For some series, in particular those in higher dimension, we obtain
satisfactory results in this manner.  Data points in the pole-residue
plots are high in number and well concentrated along a distinct line
for each series, examples of which follow below.  But for other
series, the \dlog\ method, even in combination with an
Euler-transformation, is insufficient for a quantitative
analysis.  So our strategy is in general to use the \dlog\ method only
to get rough estimates for the critical parameters, as a starting point
for a detailed analysis with M1 and M2, and to assess the general
behavior of the series from the number of pole-residue pairs which are
obtained.

\subsection[Estimation of $\pq$ and the Critical Exponents Using M1
and M2]{Estimation of $\pq$ and the Critical Exponents Using M1 and M2}

The analysis algorithms M1 and M2 allow the accurate simultaneous
determination of the threshold $\pq$, the leading critical exponent
$\gamma$, and the confluent correction to scaling
exponent $\Delta_1$, assuming the asymptotic form
\begin{equation}
  \label{eq:scaling-form-gen}
  \chi(x)\sim A(\pq - x)^{-\gamma}(1+B(\pq - x)^{\Delta_1}).
\end{equation}
In M1, one studies the logarithmic derivative of
\begin{equation}
  F(x)=\gamma \chi(x)-(x_\mathrm{c}-x)\frac{d\chi(x)}{dx}
\end{equation}
which has a pole at $x_\mathrm{c}$ with residue $-\gamma+\Delta_1$.
For a given trial value of $x_\mathrm{c}$ one obtains graphs of
$\Delta_1$ versus $\gamma$ for all Pad\'e approximants of $F$, and
chooses the triplet $x_\mathrm{c},\gamma,\Delta_1$ for which best
convergence of the different approximants results~\cite{adle93}.

In the M2 method one first transforms the series in $x$ into series in
the variable $y=1-(1-x/x_\mathrm{c})^{\Delta_1}$ and then takes Pad\'e
approximants to
\begin{equation}
G(y)=\Delta_1 (y-1)\frac{d\ln \chi}{dy}
\end{equation}
which should converge to $-\gamma$. Here one plots graphs of $\gamma$
versus the input $\Delta_1$ for different trial values of
$x_\mathrm{c}$ and again chooses the triplet $x_\mathrm{c}, \gamma,
\Delta_1$ with the best convergence of all Pad\'e approximants.  For
both methods it is advisable to perform first the usual \dlog\
analysis, to get rough estimates of $x_\mathrm{c}$ and $\gamma$ which
one uses as starting points for the detailed analysis with M1 and M2.
 The effectiveness and preciseness of
these series analysis methods has been demonstrated in several
papers~\cite{adle90,AdlerMP82,AdlerS92,GofmanAAHS93}.

In M1 we vary the trial-$\pq$ until the curves from the high order
Pad\'e approximants enter fairly symmetrically from both sides and the
best convergence is obtained. This $\pq$ and the corresponding
$\gamma$ are taken as the temporary best estimates for that series,
with temporary error estimates from the nearest trial-$\pq$'s, whose
plots show poorer convergence.  In many cases M1 proves to be quite
sensitive to small changes in the trial-$\pq$, and the degree of
convergence usually looks very convincing. Away from the best $\pq$,
convergence degrades quickly, the picture becomes non-symmetric and at
the same time the area of convergence shifts to lower or higher values
of $\gamma$.  We show examples of such plots in
Sec.~\ref{sec:explresults}.  In M2 we vary $\pq$ and look for best
convergence of the Pad\'e approximant curves while they cross each
other with a small negative slope. Compared to M1, the M2-plots are
often much less decisive.  A good convergence region sustains over a
wider range, where again the change in $\pq$ is accompanied by a shift
in the corresponding $\gamma$.

In the end we determine an overall estimate for $\pq$, which is
consistent with the estimates from both M1 and M2.  These numerical
results are presented in the tables of Sec.~\ref{sec:explresults}.  In
the tables we also include rough estimates for $\Delta_1$.  We comment
that the Euler transformation is known to produce analytic correction
terms even if not present originally.  When the leading correction
exponent is larger than $1$, as seems to be the case for some of our
series, these `artificial' corrections will show up in M1 and M2
\cite{adle84}, and hence our $\Delta_1$ estimates are mainly included
for reference and should not be trusted as the real physical values.

\subsection{Sensitivity to the Parameter of the Euler Transformation}

Our analysis relies in a large part on the use of Euler
transformations to increase the number of useful Pad\'e approximants
and to improve their convergence. The technique is well established
and has been used with success \cite{pear78}, but nevertheless we find
it worthwhile to check, to what degree our results are sensitive to
the precise choice of the parameter $\pn$, the value of $x$ that is
mapped to infinity by the transformation.  We first choose $\pn$ very
close to the negative singularity, as indicated by the \dlog\ analysis
of the original series. We then vary this $\pn$ over a considerable
range of typically 20\%, and compare the results.  We observe that a
variation of $\pn$ {\em does} move the data points or curves obtained
from individual Pad\'e approximants, but that the average (in \dlog\ 
plots) and the convergence region (in M1 plots) stay fixed to a very
good accuracy, when compared to the error bounds given by the analysis
technique itself. We thus exclude that our results are artifacts of
the applied Euler transformations.

\renewcommand{\timestamp}{Time-stamp: "2004-08-07 20:03:49 daboul"}
\newcommand{\plot}[1]{\rotatebox{90}{\includegraphics[width=.40\textheight]{sg-eps/#1}}}
\newcommand{\plotu}[1]{\rotatebox{0}{\includegraphics[height=.40\textheight]{sg-eps/#1}}}

\subsection{Explicit Results from the Analysis}\label{sec:explresults}

\sts The numerical results for dimension 8 are summarized in
Tab.~\ref{tab:dim8}.  In this dimension, even without an Euler
transformation, the \dlog\ analysis gives convincing results for all
the distributions: Bimodal, double-triangular, uniform and Gaussian.

\begin{figure}
  \caption{Position of the Poles from all Pad\'e
    approximants to the logarithmic derivative of the untransformed
    series of $\EA$ for the uniform distribution in $d=8$. The upper
    half shows the location of the poles in the complex plane. The
    lower part is a histogram of the poles along the real axis, in
    which the two peaks are of relevance. The one on the positive axis
    represents the physical singularity we want to characterize, the
    second one can interfere with analysis and may be mapped away
    using an Euler transformation.} \label{fig:uni8d}
  \centerline{\plot{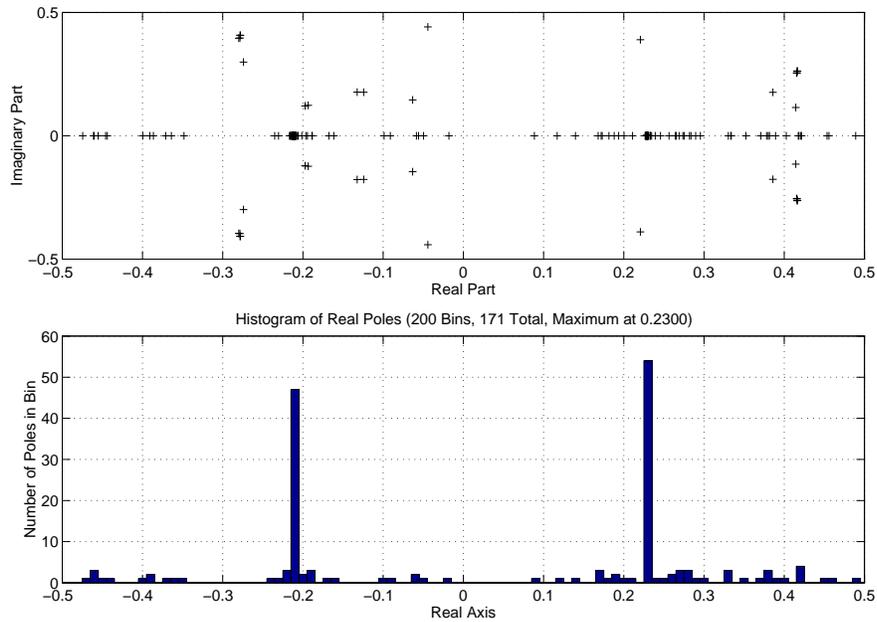}}
\end{figure}

During the analysis we prepared a large number of plots of which we
can only present a few to illustrate the process. The distribution of
the pole-estimates indicates a negative real pole for the asymptotic
function, somewhat weaker than the positive one (e.g.\
Fig.~\ref{fig:uni8d}). Both poles are at the same distance from the
origin. The convergence of pole-residue pairs improves upon Euler
transformation.

\begin{figure}
  \caption{Pole-residue plot from a Dlog Pad\'e analysis of the $\EA$-series
    for the uniform distribution in $d=8$. We use Euler
    transformations with several values $p_n$.  The main part gives an
    overview over a large region including almost all data points. The
    histograms on the axes show how these points are distributed. The
    inset is an enlarged view of the small region with the highest
    concentration of points as indicated by the box.  For comparison
    we calculate the average and standard deviation $\sigma_n$ from
    the points in the boxed area.}
  \label{fig:uni8d2}
  \centerline{\plot{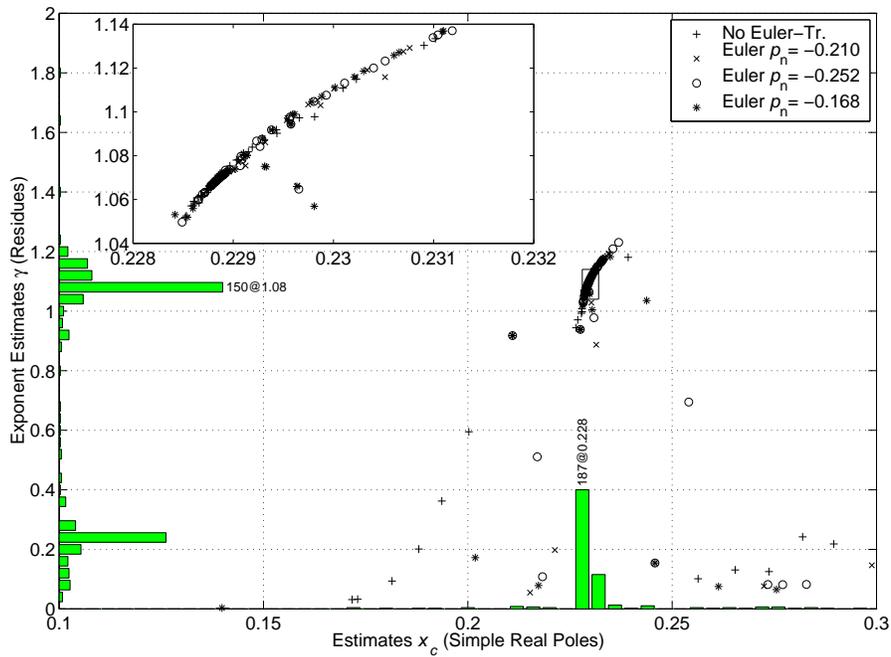}}
\end{figure}

\begin{figure}
  \caption{Estimates for the leading critical exponent
    $\gamma$ from the $\EA$-series for the double-triangular
    distribution in $d=8$. Series transformed with different values
    for $\pn$, indicated by different symbols, contribute to the plot.
    Each point corresponds to a particular transformation and Pad\'e
    approximant [L/M], where L and M relate to the number on the
    $x$-axis as given in the label.  The different symbols refer to
    different parameters for the Euler transformation: Without ($+$),
    $x_n=-0.140$ ($\times$), $x_n= -0.168$ ($\circ$), and $x_n=-0.112$
    ($*$).  Estimates are calculated as averages over points in the
    convergence region.  The points that are included in the average
    are those within the boxed area. The resulting numbers, in this
    case of $\gamma = 1.07\pm 0.02$, enter table~\ref{tab:dim8}.}
  \label{fig:tri8d}
  \centerline{\plot{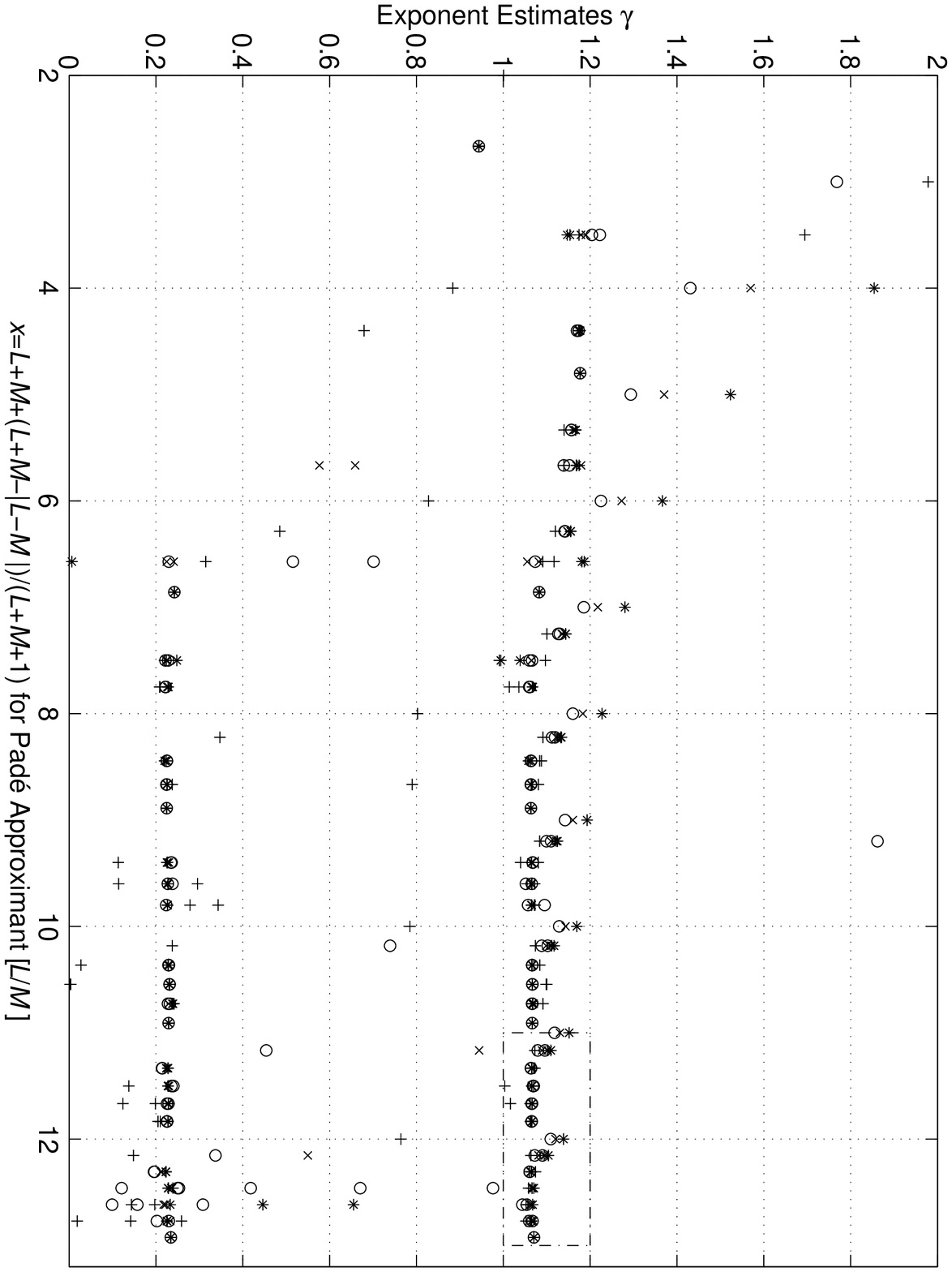}}
\end{figure}

The numerical results for the critical exponents are calculated as
averages over estimates from high-order Pad\'e approximants. These
include data from the untransformed series and from the series
transformed with three different values of $\pn$.  In the tables these
are the entries without an estimate for $\Delta_1$. All values are
slightly larger than the mean field value of $\gamma=1$
(Fig.~\ref{fig:uni8d2}).  This deviation is understood on theoretical
grounds as being due to corrections to the leading singular behavior.
Also the data indicate that the exponent estimate may further approach
1 for longer series, since generally residues decrease in value as the
approximant-order increases while remaining greater than one
(Fig.~\ref{fig:tri8d}).  When taking into account the correction
exponent with M1/M2 the deviation also decreases.  The results from M1
and M2 are shown in the table separately for different values of
$\pn$, and generally show very good agreement.  We observe the
possibility of a systematic shift to $\gamma=1.060(12)$ in case one
chooses a different region of best convergence.  The values for
$\Delta_1$ are included for reference only, due to the reason
mentioned before.  Although the absolute value of the
exponent-estimate is larger than 1, we find numerical agreement of the
results for all the tested distributions.

\begin{table}[htbp]
\caption{Results for dimension $d=8$ from the analysis with \dlog, M1
  and M2. The first line for each distribution shows the
  result from the \dlog\  analysis in which Euler
  transformations with different values $\pn$ were
  used. The remaining lines show the results from M1 in
  combination with M2, separately for several values of $\pn$.}
\label{tab:dim8}
\centerline{\begin{tabular}[c]{|l|llll|}
\hline
Distribution &Parameter &Threshold &Exponent &Correction-\\
{}           &{$\pn$}   &{}$\pc$   &$\gamma$ &Exponent \\ %
\hline
\hline
\multirow{4}{*}{Bimodal}
             &several  &0.072      & 1.05(1)  & n/a \\
             &$-0.084$ &0.07331(3) & 1.046(9) & 1.4-1.7 \\
             &$-0.073$ &0.07331(3) & 1.046(9) & 1.4-1.7 \\
             &$-0.056$ &0.07332(3) & 1.047(12) & 1.3-1.5 \\
\hline
\multirow{4}{*}{Gaussian}
             &several  &0.080      & 1.068(20)  & n/a \\
             &$-0.084$ &0.08030(3) & 1.048(9) & 1.3-1.5 \\
             &$-0.070$ &0.08029(3) & 1.047(12) & 1.3-1.5 \\
             &$-0.056$ &0.08030(3) & 1.048(9) & 1.3-1.5 \\
\hline
{}           &several  &0.148      & 1.072(22)  & n/a \\
Double-      &$-0.168$ &0.14895(3) & 1.048(9) & 1.3-1.7 \\
Triangular   &$-0.140$ &0.14895(3) & 1.048(9) & 1.3-1.5 \\
{}           &$-0.112$ &0.14898(9) & 1.048(9) & 1.3-1.5 \\
\hline
\multirow{4}{*}{Uniform}
             &several  &0.228      & 1.069(25)  & n/a \\
             &$-0.252$ &0.22852(9) & 1.048(6) & 1.3-1.5 \\
             &$-0.210$ &0.22848(6) & 1.048(6) & 1.3-1.5 \\
             &$-0.168$ &0.22854(9) & 1.048(6) & 1.3-1.5 \\
\hline
\end{tabular}}
\end{table}

\begin{figure}
  \caption{Pole-residue plot from a Dlog Pad\'e analysis of the
    untransformed $\EA$-series for the Uniform distribution in $d=7$.
    See the caption of Fig.~\ref{fig:uni8d2} for further explanation.}
  \label{fig:uni7d}
  \centerline{\plot{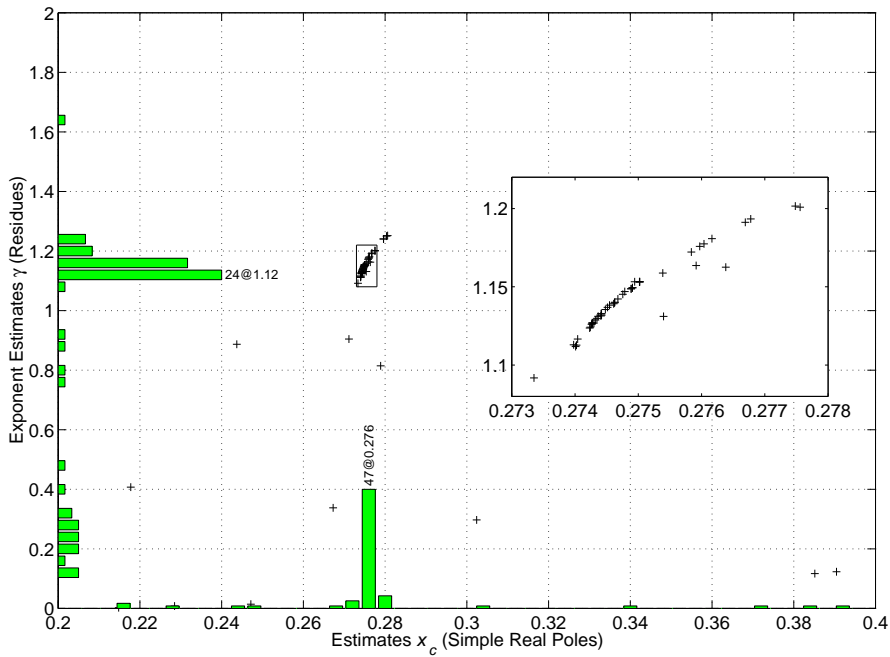}}
\end{figure}

The qualitative behavior in dimension 7 agrees with that for $d=8$, although
the exponent estimates are slightly farther away from 1 (see
Tab.~\ref{tab:dim7}).  The critical threshold $\pc$ for each series is
larger than in dimension 8.  Again we observe a negative pole of
comparable strength and distance from the origin as the physical one,
but nevertheless, even without an Euler transformation the series give
consistent results.  To illustrate this we show in
Fig.~\ref{fig:uni7d} a pole-residue plot of the untransformed series
for the uniform distribution and also include in Tab.~\ref{tab:dim7}
the estimates from the \dlog\ analysis without transformation. In our
M1/M2-analysis we again observe the possibility of a systematic shift
to $\gamma=1.120(15)$ in case one chooses a different region of best
convergence.

\begin{table}[tbp]
  \caption{Results for dimension $d=7$ from the analysis with \dlog, M1
    and M2. The first line for each distribution shows the result
    from the \dlog\ analysis in which no Euler transformations was
    applied.  The remaining lines show the results from M1 in
    combination with M2, separately for several values of
    $\pn$.}\label{tab:dim7}
  \centerline{\begin{tabular}[c]{|l|llll|}
      \hline
Distribution &Parameter &Threshold &Exponent &Correction-\\
{}           &{$\pn$}   &{}$\pc$   &$\gamma$ &Exponent \\ %
\hline
\hline
\multirow{4}{*}{Bimodal}
             &none     &0.088      & 1.14(3)   & n/a \\
             &$-0.082$ &0.08731(9) & 1.105(15) & 1.5-1.7 \\
             &$-0.078$ &0.08732(9) & 1.105(12) & 1.3-1.7 \\
             &$-0.051$ &0.08738(6) & 1.110(15) & 1.3-1.4 \\
\hline
\multirow{3}{*}{Gaussian}
             &none     &0.097(1)   & 1.14(3)   & n/a \\
             &$-0.071$ &0.09710(6) & 1.107(9)  &1.3-1.4 \\
             &$-0.062$ &0.09712(9) & 1.108(12) & $\approx$1.3 \\
\hline
{}           &none     &0.1784(4)  & 1.14(3)   & n/a \\
Double-      &$-0.265$ &0.17790(9) & 1.108(9) & 1.3-1.6 \\
Triangular   &$-0.156$ &0.17790(9) & 1.108(9) & 1.3-1.5 \\
{}           &$-0.112$ &0.17799(9) & 1.110(6) & $\approx$1.3 \\
\hline
\multirow{4}{*}{Uniform}
             &none     &0.2745(5)  & 1.13(3)   & n/a \\
             &$-0.315$ &0.27402(9) & 1.108(9) & 1.3-1.5 \\
             &$-0.234$ &0.27399(9) & 1.108(6) & 1.3-1.5 \\
             &$-0.225$ &0.27399(9) & 1.108(6) & 1.3-1.5 \\
\hline
\end{tabular}}
\end{table}

\begin{figure}
  \caption{Pole-residue plot from a Dlog Pad\'e
    analysis of the $\EA$-series for the Gaussian distribution in
    $d=5$. We use Euler transformations with several values $p_n$.
    See the caption of Fig.~\ref{fig:uni8d2} for further
    explanation.} \label{fig:gau5d1}
  \centerline{\plot{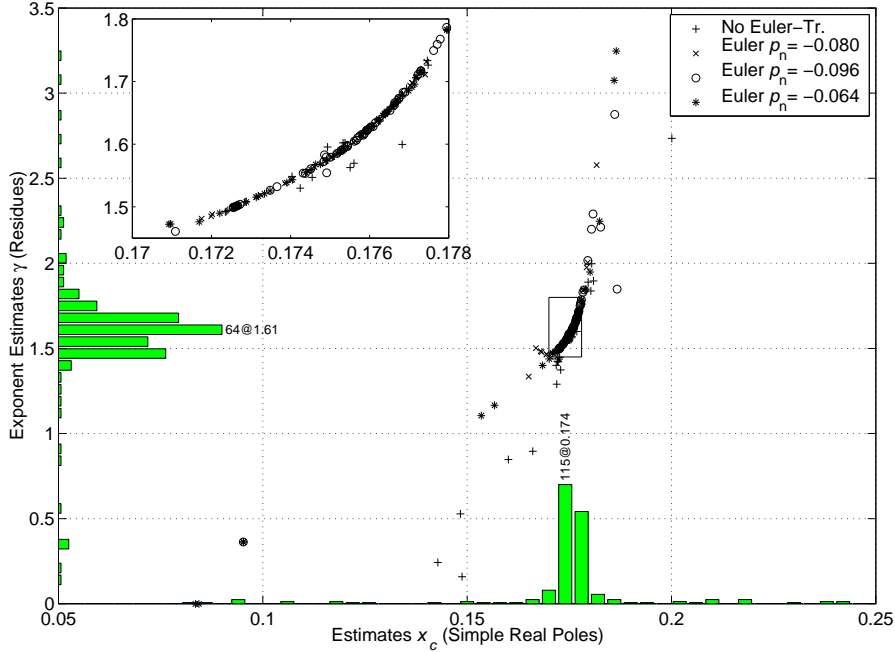}}
\end{figure}

\begin{figure}
  \caption{Estimates for the leading critical exponent
    $\gamma$ from the $\EA$-series for the Gaussian distribution in
    $d=5$. See the caption of Fig.~\ref{fig:tri8d} for details.  The
    distribution of the data-points indicates that the
    $\gamma$-estimate may still increase with longer series.  The
    different symbols refer to different parameters for the Euler
    transformation: Without ($+$), $x_n=-0.080$ ($\times$), $x_n=
    -0.096$ ($\circ$), and $x_n=-0.064$ ($*$). Averaging over the
    points inside the small box leads to an estimante of $\gamma =
    1.68\pm 0.08$, which enters table~\ref{tab:dim5}. }
  \label{fig:gau5d2}
  \centerline{\plot{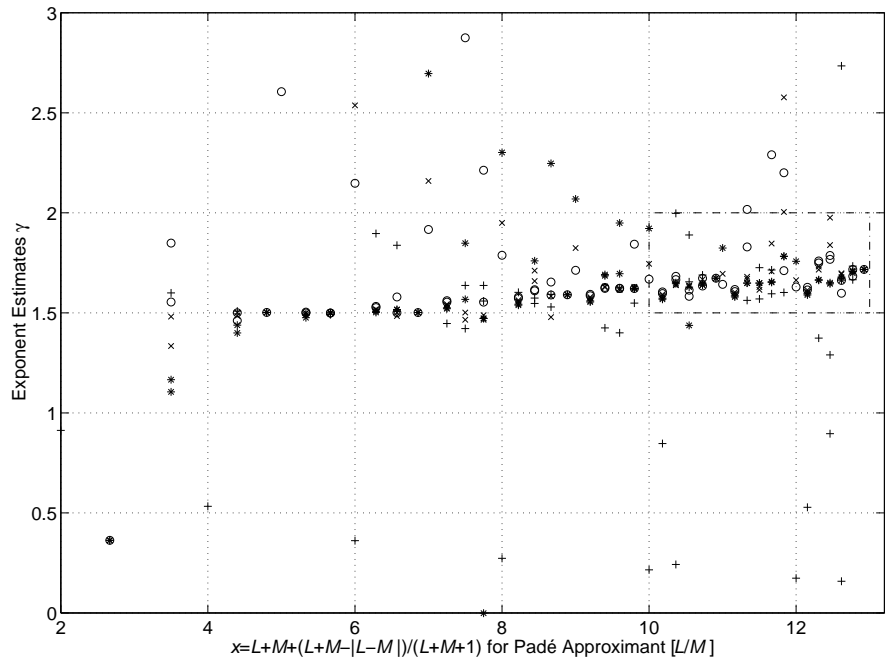}}
\end{figure}

In dimension 5 the negative pole is closer to the origin than the
positive one, and convergence degrades. Still the data points line up
properly.  To improve their convergence and to get higher numerical
accuracy we apply Euler transformations with several $\pn$ (e.g.\
Fig.~\ref{fig:gau5d1}) and the final estimates (Tab.~\ref{tab:dim5})
are obtained the same way as described before.  Looking at
Fig.~\ref{fig:gau5d2}, the main line of data-points for the Gaussian
distribution still increases with the order of the Pad\'e
approximants. Therefore, the value measured by this method is probably smaller
than what a longer series would show.  In summary, all studied
distributions agree at $d=5$ on a common exponent within their error
margins.

\begin{figure}
  \caption{M1 analysis of the $\EA$-series for the Gaussian
    distribution in $d=5$. An Euler transformation with $\pn=0.08$ was
    applied. Plot (a) was obtained for a trial $x$ of 0.177, where
    convergence is visible but lies below the optimal convergence
    point. In (b) the trial $x=0.179$ is near or at the point of
    optimal convergence.}\label{fig:gau5dM1_1}
  \begin{center}
    (a)\hfill\\
    \plotu{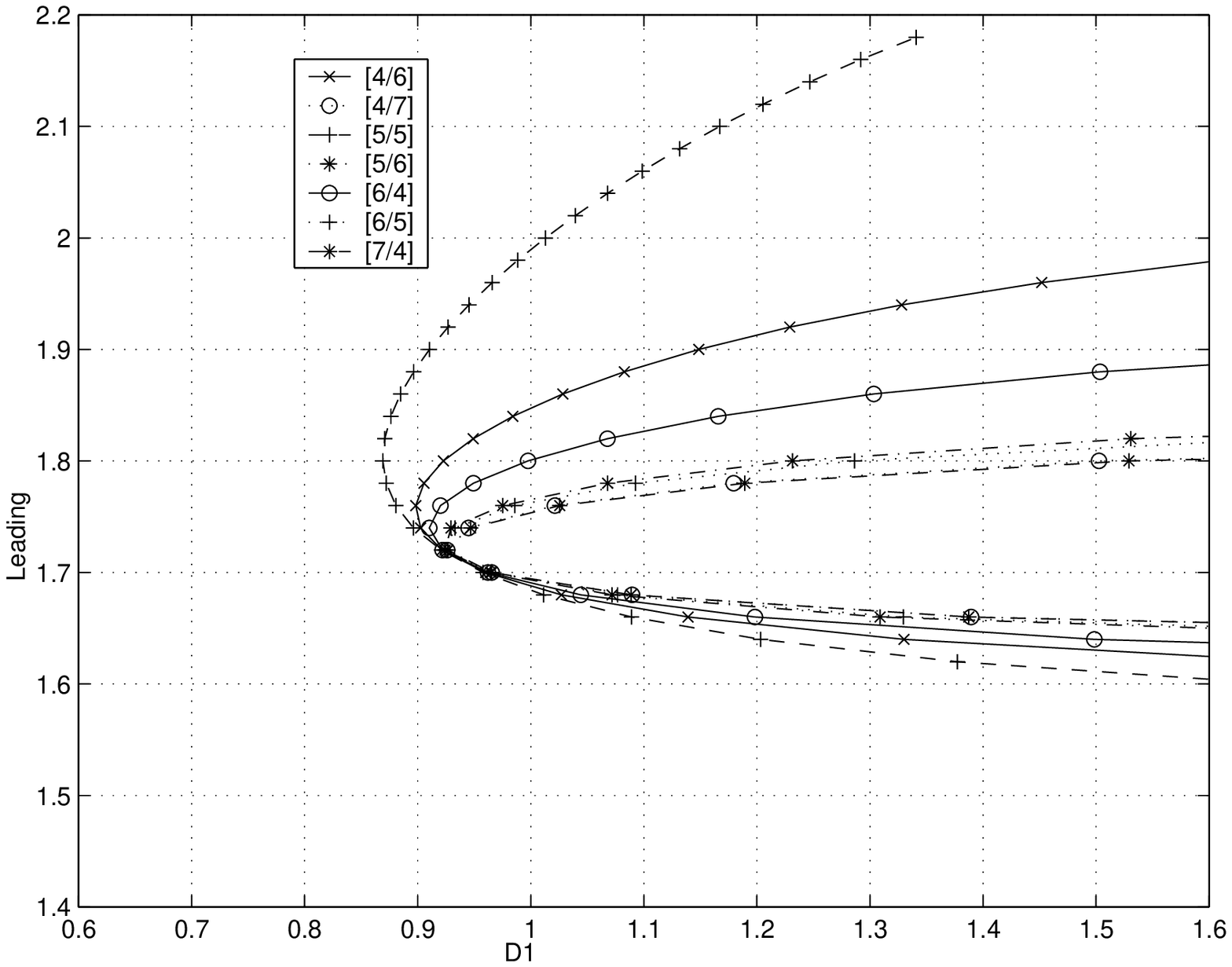}\\
    (b)\hfill\\
    \plotu{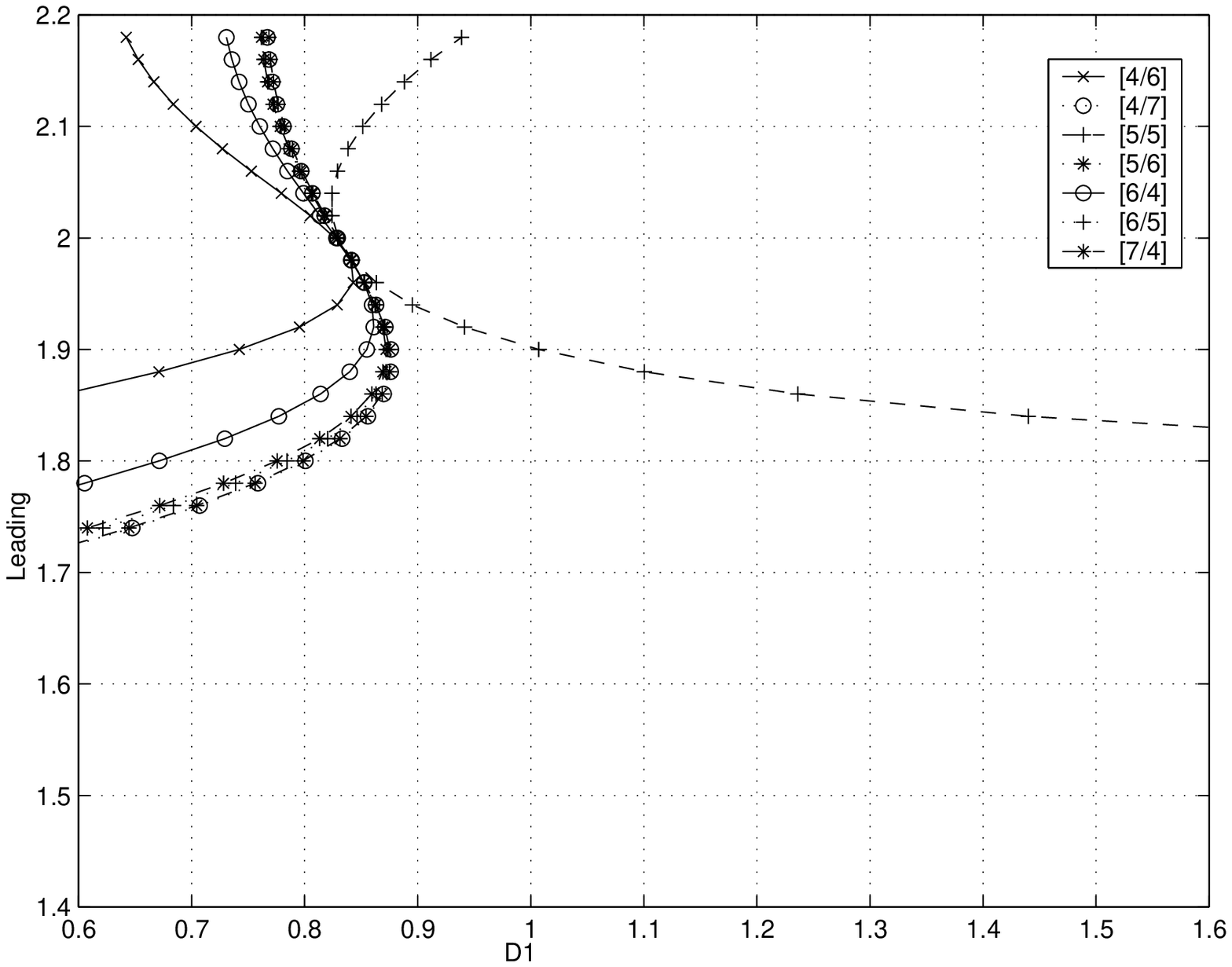}
  \end{center}
\end{figure}

\begin{figure}
  \begin{center} \caption{M1 analysis of the untransformed
    $\EA$-series for the Gaussian distribution in $d=5$. The trial
    $x=0.179$ is near or at the point of optimal convergence. Compared
    to Fig.\ref{fig:gau5dM1_1}(b), with Euler-transformation,
    convergence is less sharp and shifted to a slightly larger
    $\gamma$. Strong fluctuations of the Pad\'e-lines are visible in the
    lower part of the plot.}\label{fig:gau5dM1_0}

    \plotu{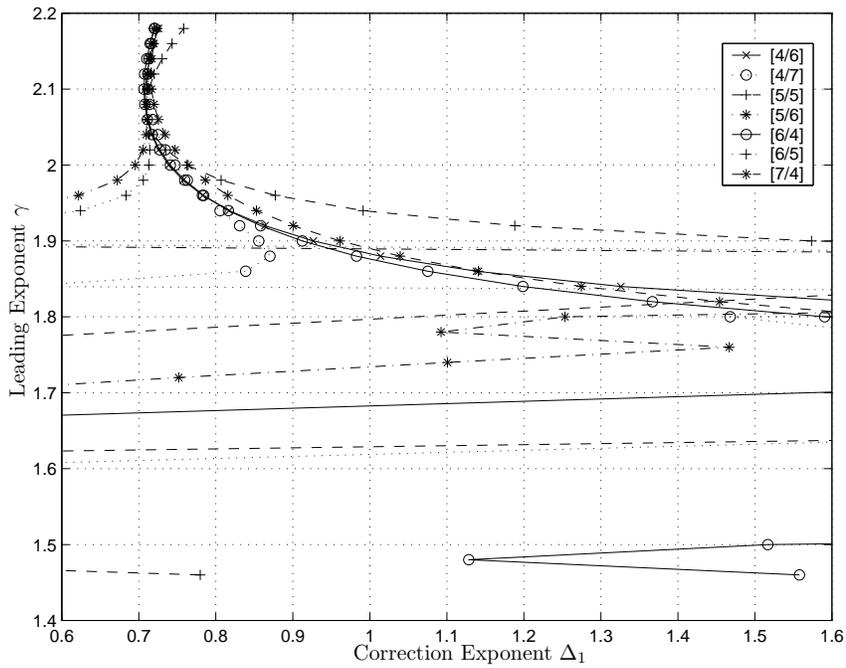}
  \end{center}
\end{figure}

\begin{figure}
  \begin{center} \caption{M2 analysis of the Euler-transformed
    $\EA$-series ($\pn=0.08$) for the Gaussian distribution in $d=5$
    and for a trial $x$ of 0.179. Illustrates the need to use M1 and
    M2 in combination.}\label{fig:gau5dM2_1}

    \plotu{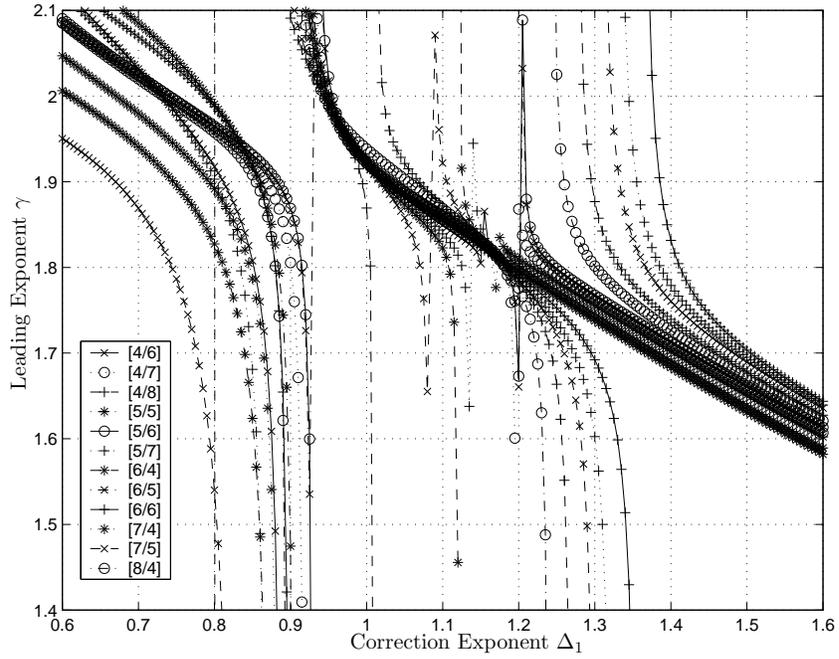}
  \end{center}
\end{figure}

\begin{figure}
  \begin{center} \caption{M1 analysis of the $\EA$-series for the
    Uniform distribution in $d=5$, after an Euler-transformation with
    $\pn=0.29$. $x=0.488$ is near or at the the point of optimal
    convergence.}\label{fig:uni5dM1_1}

    \plotu{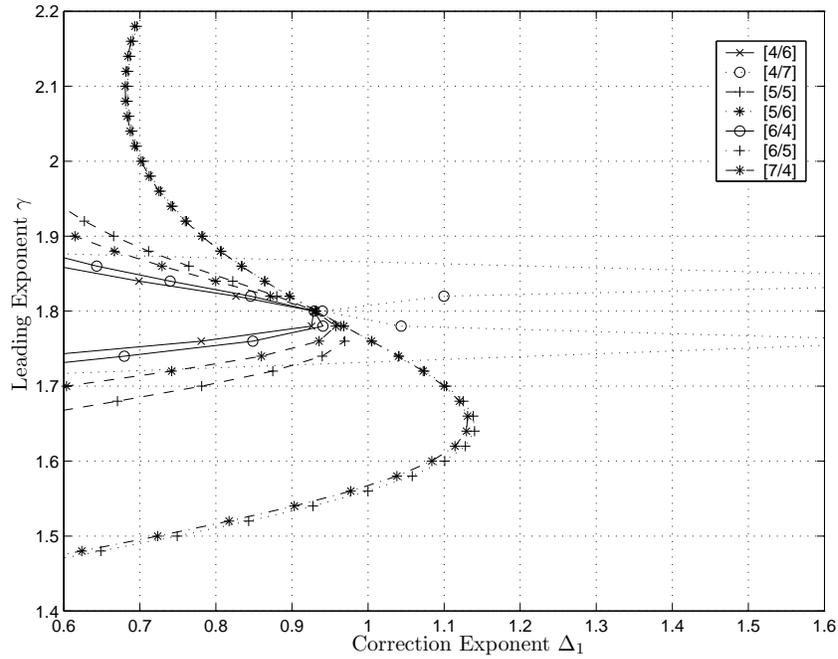}
  \end{center}
\end{figure}

In Figures~\ref{fig:gau5dM1_1} to \ref{fig:gau5dM1_0} we show, for the
case of the Gaussian distribution, plots as they are typically
obtained from M1.  Each curve in a plot comes from a different Pad\'e
approximant as the legend shows.  In all figures we clearly see a
region where the lines converge, and since all figures show the same
range in the $\gamma$-$\Delta_1$-plane it is also easy to see that the
convergence region shifts around.  Figures~\ref{fig:gau5dM1_1}(a) and
(b) differ in the trial $x$ as input parameter.  While
for $x=0.177$ convergence is quite good, is becomes better for
$x=0.179$ (in fact best among our trial values).  At this value of $x$
the shape of the curves is also symmetric and they switch over to one
side for larger $x$ (the opposite side, when compared to the smaller $x$)
which we find to be a characteristic feature of the point of best
convergence.  Figure~\ref{fig:gau5dM1_0} shows the corresponding plot
for the untransformed series. Here $x=0.179$ is also near the
characteristic point of symmetric curves.  Still convergence is not as
good as in Fig.~\ref{fig:gau5dM1_1}(b) and the center is shifted to a
slightly larger value of $\gamma$.  The example, although not the most
common case, also illustrates that one must not rely on analysis with
either M1 or M2 alone.  While M1 gives an estimate of $\gamma=2$ or
higher, M2 (Fig.~\ref{fig:gau5dM2_1}) points to a lower value of
roughly $1.82$ and $\Delta_1$ above 1, or $\gamma\approx 1.95$ with
$\Delta_1$ below 1 and poorer convergence.  Our estimates in
the tables always result from using M1 together with M2.
Fig.~\ref{fig:uni5dM1_1} is another example, showing a plot from
M1-analysis for the Uniform distribution near the symmetry point of
best convergence.

\begin{table}[tbp]
\caption{Results for dimension $d=5$ from the analysis with \dlog, M1 and
  M2. See caption of Tab.~\ref{tab:dim8} for details.}
\label{tab:dim5}
\centerline{\begin{tabular}[c]{|l|llll|}
\hline
Distribution &Parameter &Threshold &Exponent &Correction-\\
{}           &{$\pn$}   &{}$\pc$   &$\gamma$ &Exponent \\ %
\hline
\hline
\multirow{3}{*}{Bimodal}
             &several  &0.154    & 1.91(10) & n/a\\
             &$-0.120$ &0.154(3) & 1.95(15) & 1.1-1.3 \\
             &$-0.100$ &0.154(3) & 1.95(15) & $\approx$1.0 \\
\hline
\multirow{4}{*}{Gaussian}
             &several  &0.174    & (1.67(8)) & n/a\\
             &$-0.096$ &0.176(3) & 1.70(15) & 0.8-1.0 \\
             &$-0.080$ &0.177(3) & 1.75(15) & 0.8-1.0 \\
             &$-0.064$ &0.177(3) & 1.75(15) & 0.8-1.0 \\
\hline
\multirow{3}{*}{\begin{minipage}{11ex}Double-\\ Triangular\end{minipage}}
             &several  &0.312    & 1.81(7)& n/a\\
             &$-0.240$ &0.312(6) & 1.80(15) & 0.9-1.0 \\
             &$-0.200$ &0.312(6) & 1.80(15) & 0.9-1.0 \\
\hline
\multirow{3}{*}{Uniform}
             &several  &0.484    & 1.72(6)& n/a\\
             &$-0.348$ &0.484(6) & 1.70(15) & 1.0-1.2 \\
             &$-0.290$ &0.487(6) & 1.70(15) & 0.8-1.0 \\
\hline
\end{tabular}}
\end{table}

\psfrag{Leading}[bl][bl][0.75]{Leading Exponent $\gamma$}
\psfrag{D1}[tl][tl][0.75]{Correction Exponent $\Delta_1$}

As we decrease the dimension further to $d=4$, the analysis becomes
increasingly difficult. One reason is the negative pole on the real
axis, which is very strong and apparent for all the series.  Without
an Euler transformation, the \dlog\ analysis does not show anything
conclusive. For the transformed series, a larger number of data points
lines up well in the pole-residue plots, but they are still not well
converged.  The series for the Gaussian distribution is somewhat
exceptional here: Exponent estimates converge well with increasing
approximant-order, but they approach a value of $\gamma=3.1\pm0.1$
which is much higher than what we obtain in the other cases and with
the other methods.  If this is not simply an artifact of the
transformation, it must be attributable to the correction-term to
scaling, which becomes increasingly important at lower dimension.

With M1/M2 applied to the transformed series we are able to obtain
estimates for $\pc$ and $\gamma$, although the error margins are quite
large.  Indeed, the value we obtain for the correction exponent
$\Delta_1$ is much larger than $1$ (and larger than in the higher
dimensions). Again we find that our numbers agree with a common
exponent $\gamma$ for all the distribution functions.

\begin{table}[htbp]
\caption{Results for dimension $d=4$ from the analysis with \dlog, M1
  and M2. The table lines without estimates for $\Delta_1$ stem
  from the \dlog\  analysis in which Euler
  transformations with different values $\pn$ were
  used. The remaining lines show the results from M1 in
  combination with M2, separately for several values of $\pn$.}
\label{tab:dim4}
\centerline{\begin{tabular}[c]{|l|llll|}
\hline
Distribution &Parameter &Threshold &Exponent &Correction-\\
{}           &{$\pn$}   &$\pc$   &$\gamma$ &Exponent \\ %
\hline
\hline
\multirow{2}{*}{Bimodal}
             &$-0.144$ &0.26(2) & 2.5(3) & 1.5-1.6 \\
             &$-0.120$ &0.26(2) & 2.5(3) & 1.5-1.6 \\
\hline
\multirow{4}{*}{Gaussian}
             &several  &0.31(2) &  3.1(1) & n/a\\
             &$-0.108$ &0.312(4) & 2.3(1) & 1.3-1.4 \\
             &$-0.090$ &0.314(4) & 2.3(1) & 1.3-1.4 \\
             &$-0.072$ &0.314(4) & 2.3(1) & 1.3-1.4 \\
\hline
\multirow{3}{*}{\begin{minipage}{11ex}Double-\\ Triangular\end{minipage}}
             &several  &0.52(8) & 2.8(8) & n/a\\
             &$-0.276$ &0.54(2) & 2.5(2) & $\approx$1.5 \\
             &$-0.230$ &0.54(2) & 2.5(2) & $\approx$1.5 \\
\hline
\multirow{3}{*}{Uniform}
             &$-0.396$ &0.84(2) & 2.5(1) & 1.3-1.4 \\
             &$-0.330$ &0.83(2) & 2.4(1) & 1.3-1.4 \\
             &$-0.264$ &0.84(2) & 2.4(1) & 1.3-1.4 \\
\hline
\end{tabular}}
\end{table}

\renewcommand{\timestamp}{Time-stamp: "2004-05-03 09:32:39 daboul"}
\section{Conclusions\label{sec:conclusion}}

\begin{figure}
  \begin{center}
    \caption{Estimates for the critical exponent $\gamma$ grouped by
      space dimension. The different values for each dimension are
      obtained from the 4 probability distributions. The estimates
      appear to agree on a common exponent $\gamma$ within but
      separately in each dimension. Above the critical dimension
      $d_u=6$ the values are close to 1.}
    \label{fig:conclusion}
    \includegraphics[height=.25\textheight]{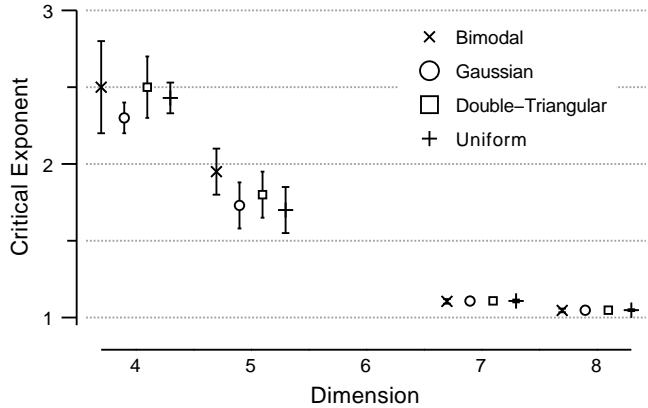}
  \end{center}
\end{figure}

\sts Figure~\ref{fig:conclusion} summarizes our
numerical estimates for the leading critical exponent
$\gamma$ of the Edwards Anderson susceptibility $\EA$ in the different
dimensions.  For each dimension we show the 4 values obtained for the
different random distributions.  The mean values and error bars are as
shown in Tables~\ref{tab:dim8} to \inthesis{4.8.}{8.}

The error bars for dimensions 7 and 8 are too small to be visible
in the plot.  We observe once more that in these dimensions the
estimates are close to, but still larger than the expected mean-field
value of 1.  As we argued, the deviation is likely caused by
correction to scaling terms and the fact that we work with relatively
short series.  Our \dlog\ analysis suggests smaller estimates with
longer series, and when accounting for the first correction term using
the methods M1 and M2 the deviation indeed decreases, but higher order
corrections cannot be excluded.  The larger deviation for $d=7$ is
consistent with the observation that the correction terms are more
dominant in lower dimensions.

As expected, the estimates we obtain for dimensions 4 and 5 are pronouncedly
different from the mean field value.  Within each dimension the
estimates agree on a common value for all the random distributions we
study, which is roughly $\gamma=2.4\pm0.2$ in $d=4$ and
$\gamma=1.8\pm0.2$ in $d=5$.
Thus our data do not indicate that the random distribution for the
quenched-in disorder splits the spin glass model into many
universality classes nor that the model behaves in that respect
differently than other common thermodynamic models, in contrast to
claims from Refs.~\cite{camp94e, camp94b, BernardiC95, BernardiPC96,
  BernardiC97, BernardiLMCAC98, CampbellPMB00}.  Instead we find
confirmation for the established picture, that the space dimension
creates universality classes and that the leading critical exponent is
a universal quantity~\cite{BKpriv91}.

Most of the simulations of the Ising spin glass have been done in
dimension 3, in which our series do not perform well. For a direct
comparison we are thus limited to the sparse results for $d=4$
from~\cite{BernardiC95,BernardiC97}. Our estimates for the critical
temperature $T_c$ agree rather well with those by Bernardi and
Campbell.  The compared values are
$T_c=1.96$\ (vs.\ $1.99\pm0.01$) for the bimodal distribution,
$T_c=1.88$\ (vs.\ $1.91\pm0.01$) for the uniform distribution, and
$T_c=1.79$\ (vs.\ $1.77\pm0.01$) for the Gaussian distribution, where
the uncertainty in our estimates is also roughly 1 in the last digit,
from fluctuations and from possible additional systematic shifts, due to scaling
corrections.  We confirm a slight decrease of $T_c$ with
increasing kurtosis of the random distributions, which is defined as
the ratio of moments $R=M_4/M_2^2$.
The kurtosis values are: Bimodal 1,
double-triangular 4/3, uniform 9/5, Gaussian 3 and for the exponential
distribution 6. Bernardi and Campbell have calculated the exponent
$\eta$, while we have values for $\gamma$, so we currently lack a
third exponent, such as $\nu$, for a direct comparison. However, the
discrepancy in the general universal vs.\ non-universal behavior
remains.

Some authors~\cite{KawashimaY96, BallesterosCFMPRTTUU00, MariC02} have stressed the importance of
corrections to finite size scaling (FSS). In taking these into
account, they do not find violated universality.
We cannot assess the quality of the simulations that were done by
Bernardi, Campbell~\cite{camp94e, camp94b, BernardiC95,
BernardiLMCAC98} and others,
but, from the data in the papers and later citations, we are led to
speculate that neglected corrections to FSS have caused systematic
errors in the exponent estimates from simulations.
We would further like to stress the
general statement made by other authors, that the characteristic
features of the spin glass with its quenched-in disorder creates
enormous problems for simulations.  Series expansion techniques appear
here particularly suitable since the configurational average over the
randomness is handled {\em exactly} within their framework and own
limitations.

\inthesis{}{\section*{Acknowledgments}
  We thank the German Israeli Foundation for support.}

\bibliographystyle{osa}
\bibliography{sg}
\ifthenelse{\boolean{bo:aa}}{\linespread{1.3}}{}
\end{document}